\documentclass[sigplan,10pt]{acmart}
\renewcommand\footnotetextcopyrightpermission[1]{}
\usepackage{epsfig,endnotes}
\usepackage{times}
\usepackage{algpseudocode}
\usepackage{algorithm}

\usepackage{algorithm, algorithmicx, algpseudocode}

\usepackage{url}
\usepackage{booktabs} 
\usepackage{balance}
\usepackage{float}
\usepackage{subfig}
\usepackage{graphicx}
\usepackage{listings}
\usepackage{setspace}
\usepackage{xspace}
\usepackage{color}

\usepackage{tikz}

\usepackage{enumitem}

\usepackage[font={footnotesize,it}]{caption}
\usepackage[export]{adjustbox}
\usepackage[skip=1ex]{caption}

\usepackage[small, compact]{titlesec}
\titlespacing*{\section}{1pt}{3.5pt}{2pt}
\titlespacing*{\subsection}{1pt}{3pt}{1.5pt}
\titlespacing*{\subsubsection}{1pt}{3pt}{1.5pt}

\usepackage{amssymb}
\usepackage{pifont}

\providecommand\phantomsection{}

\newcommand{\secref}[1]{{\S\ref{#1}}}

\newcommand{\arjun}[1]{}
\newcommand{\bala}[1]{}
\newcommand{\aditya}[1]{}
\newcommand{\shivaram}[1]{}
\renewcommand{\arjun}[1]{{\color{red}{\bf AS:#1}}}
\renewcommand{\bala}[1]{{\color{brown}{\bf AB:#1}}}
\renewcommand{\aditya}[1]{{\color{blue}{\bf AA:#1}}}
\renewcommand{\shivaram}[1]{{\color{orange}{\bf SV:#1}}}

\newcommand{\myvec}[1]{\protect\overrightarrow{#1}}

\newcommand{\name}{Archipelago\xspace}

\definecolor{nodecolor}{RGB}{255,115,115}

\begin{document}

\title{\huge \name: A Scalable Low-Latency Serverless Platform}

\author{Arjun Singhvi}
\affiliation{%
  \institution{University of Wisconsin-Madison}
}	

\author{Kevin Houck}
\affiliation{%
  \institution{University of Wisconsin-Madison}
}	

\author{Arjun Balasubramanian}
\affiliation{%
  \institution{University of Wisconsin-Madison}
}	

\author{Mohammed Danish Shaikh}
\affiliation{%
  \institution{University of Wisconsin-Madison}
}	

\author{Shivaram Venkataraman}
\affiliation{%
  \institution{University of Wisconsin-Madison}
}	

\author{Aditya Akella}
\affiliation{%
  \institution{University of Wisconsin-Madison}
}	

\date{}
\settopmatter{printacmref=false, printccs=false, printfolios=true}
\begin{abstract}
The increased use of micro-services to build web applications has
spurred the rapid growth of Function-as-a-Service (FaaS) or serverless
computing platforms. While FaaS simplifies provisioning and scaling for application developers, it introduces new challenges
in resource management that need to be handled by the cloud
provider. Our analysis of popular serverless workloads
indicates that schedulers need to handle functions that are very
short-lived, have unpredictable arrival patterns, and require
expensive setup of sandboxes. The challenge of running a large number
of such functions in a multi-tenant cluster makes existing scheduling
frameworks unsuitable.

We present \name{}, a platform that enables low latency
request execution in a multi-tenant serverless setting. \name{} views
each application as a DAG of functions, and every DAG in associated
with a latency deadline. \name{} achieves its per-DAG request latency goals by:
(1) partitioning a given cluster into a number of smaller worker pools, and
associating each pool with a semi-global scheduler (SGS), (2) using 
a latency-aware scheduler within each SGS along with proactive sandbox allocation to reduce
overheads, and (3) using a load balancing layer to route requests for different DAGs to the appropriate SGS,
and automatically scale the number of SGSs per DAG. Our testbed results show that \name{} meets the latency deadline
for more than 99\% of realistic application request workloads, and reduces tail latencies by up to
\textasciitilde$36X$ compared to state-of-the-art serverless
platforms.
\end{abstract}

\maketitle
\pagestyle{plain}

\section{Introduction}
\label{sec:introduction}

Recent trends in cloud computing point towards increased adoption
of micro-services to design and deploy online
applications~\cite{berkeley_report}. These micro-services are typically
designed to compute a single function with the goal that each
micro-service can be independently deployed and managed in a cluster,
and collectively the microservices implement what used to be realized
as large monolithic applications. To meet this demand imposed by
independently scalable functions, simplify programming, and relieve programmers from
provisioning and elastic scaling responsibilities,
cloud computing providers now offer Function-as-a-Service
(FaaS) or {\em serverless} computing~\cite{awslambda,AzureFaas,googlecloudfunctions} offerings, such
as, AWS Lambda, Azure Functions, Google Cloud Functions etc.

While serverless computing simplifies a number of aspects of designing
and deploying microservice workloads in the cloud, it introduces a
number of new challenges with respect to resource management and
scheduling for the cloud provider.
The specific workload properties that
make scheduling challenging, especially in a multi-tenant setting that
supports microservices from different applications, include: (i)
function execution times are typically short-lived with 90\% of
functions executing for less than a second, but a few functions
execute for 10s of seconds (\secref{sec:motivation}); (ii) as functions are expected to be isolated, they often require setting up appropriate computational units, or ``sandboxes'', but these sandboxes can be reused to serve future
function requests; and, (iii) the arrival patterns of application
requests as a whole, and for microservices or functions therein, can
vary substantially making it necessary for the scheduler to handle
large dynamic variations in the workload.

Existing architectures for scheduling and resource allocation in large
clusters are unable to handle the above requirements.  Centralized
schedulers~\cite{yarn,mesos,kubernetes} cannot scale to handle the low
latency and high requests-per-second throughput requirements, nor are
they designed to offer good performance under rapidly-changing request
arrival patterns. On the other hand, decentralized approaches (e.g.,
Sparrow~\cite{sparrow} or Ray~\cite{ray})), where multiple schedulers
with a global view carry out scheduling (e.g., by randomly probing
machines) are more scalable, but may not find machines that have a
sandbox available for reuse leading to additional overheads from
sandbox setup. Finally, existing frameworks do not account for the
execution time of individual functions and thus are unable to
appropriately prioritize DAG requests to ensure that the end-to-end
latencies, which may include sandbox provisioning and setup, are as
close as possible to the execution time for a vast majority of
incoming application requests.

We present \name{}, a scheduling framework that
supports low overhead function execution, and enables tight latencies for
application request completions in a multi-tenant serverless
setting. \name{} views each application as a DAG, where nodes are
microservices or functions, and edges are I/O dependencies, and allows
the programmer to associate a deadline with the DAG. As requests
arrive at variables rates for different DAGs, \name{} schedules the
execution of the constituent functions on a given cluster of resources
such that a vast majority of incoming requests meet their
deadline. 

\name{} achieves the above goal via a combination of
techniques. First, \name{} \emph{partitions} the given cluster into a
number of smaller worker pools. Each worker pool is managed by a
\emph{semi-global} scheduler (SGS); with appropriate sizing of the
worker pool, we can ensure that each SGS imposes low scheduling
overheads for request execution. To achieve optimal placement
and ensure that most incoming requests are served by a ready sandbox,
each SGS also tracks the number of requests sent for every DAG it is
serving, and proactively allocates sandboxes to minimize the overheads
in launching DAGs' functions. Crucially, we create these sandboxes as
\emph{soft state} where they only use memory resources from a fixed
sized pool and can be evicted without affecting correctness.

Second, \name{} uses a scheduling algorithm within an SGS that is
aware of the latency requirements for each DAG.  This enables us to
compute a running \emph{slack}, or the time remaining for a given
DAGs' request,  and use a variant of the
shortest-remaining-time-first algorithm to minimize the possibility
of DAGs missing their \emph{deadlines}. Here, we leverage the fact that
applications running in a cluster have different slacks, and low-slack
applications' resource needs can be met by reallocating resources away
from high-slack ones. 

While partitioning a cluster can help lower scheduling
overheads, we must determine how requests are routed to each
SGS in a cluster. Thus, the third idea in \name{} is to use a
sandbox-aware load balancing layer that can route requests while being
aware of the number of sandboxes of different DAGs 
allocated in every SGS. In order to simplify the design of the load
balancing layer and make it scalable, every application DAG running in
the cluster is assigned to a single SGS to begin with and based on the
number of requests, the load balancer can either \emph{scale out} (or scale
in) the number of SGS assigned to this DAG. Using an approach that is
also aware of sandbox allocation ensures that application performance
is minimally affected when scaling across the cluster.

We build \name{} in Go and evaluate our prototype against
the current state-of-the-art serverless scheduler using a collection 
of applications derived from our analysis of real-world serverless workloads.
Our results show that
\name{} is able to meet the latency deadline for more than 99\% of
requests across various application classes, and reduces tail
latencies by more than $36\times$. We find that sandbox-aware load
balancing can reduce tail latencies by up to $24.38\times$, and that
\name{}'s sandbox placement policy is crucial to meeting latency
deadlines.
\section{Background and Motivation}
\label{sec:motivation}
We start by providing a primer on serverless computing. We then characterize
the properties of real world serverless applications available on the
repository maintained by AWS~\cite{aws}. Based on our analysis, we state our requirements and end with why current
serverless platforms fall short.

\subsection{Serverless Computing Background}

In serverless computing or FaaS, the programmer develops an
application (or simply an ``app'') as a directed acyclic graph (DAG)
of functions, uploads it to the serverless platform (which stores the
code in a datastore) and registers for an event (e.g., incoming HTTP
requests, object uploads) to trigger its execution. The platform
triggers the DAG execution only when the event arrives, and thus the
programmers are billed only when the DAG runs and for the cumulative
execution times of the constituent functions. Henceforth, we use event
and request interchangeably.

Internally, the platform consists of a load
balancing layer, a scheduling layer, and cluster machines. When a request
arrives at one of the load balancers, it routes the request to one of the many
internal schedulers. The scheduler triggers the execution of the root
function(s) of the corresponding DAG by setting up sandboxes (involves
launching a new container, setting up the runtime environment, and deploying
the function by downloading the code from the datastore) on the machines in
the cluster and running the function(s).

Alternatively, the function can
directly run on a ``warmed up'' sandbox as platforms typically do not
immediately decommission sandboxes enabling reuse for future executions of the
same function. On completion, a notification is sent to the scheduler, which
then triggers the execution of the downstream functions. The process repeats
until DAG completion. Additionally, the platform elastically scales by
launching more sandboxes based on incoming events.

\subsection{Characterizing Real World Serverless Apps}
\label{subsec:real_world_study}

 \begin{figure*}[t]
 	 \captionsetup[subfloat]{captionskip=-3pt}
 	\centering 
 	\subfloat[][]{%
 		\includegraphics[width=0.25\textwidth]{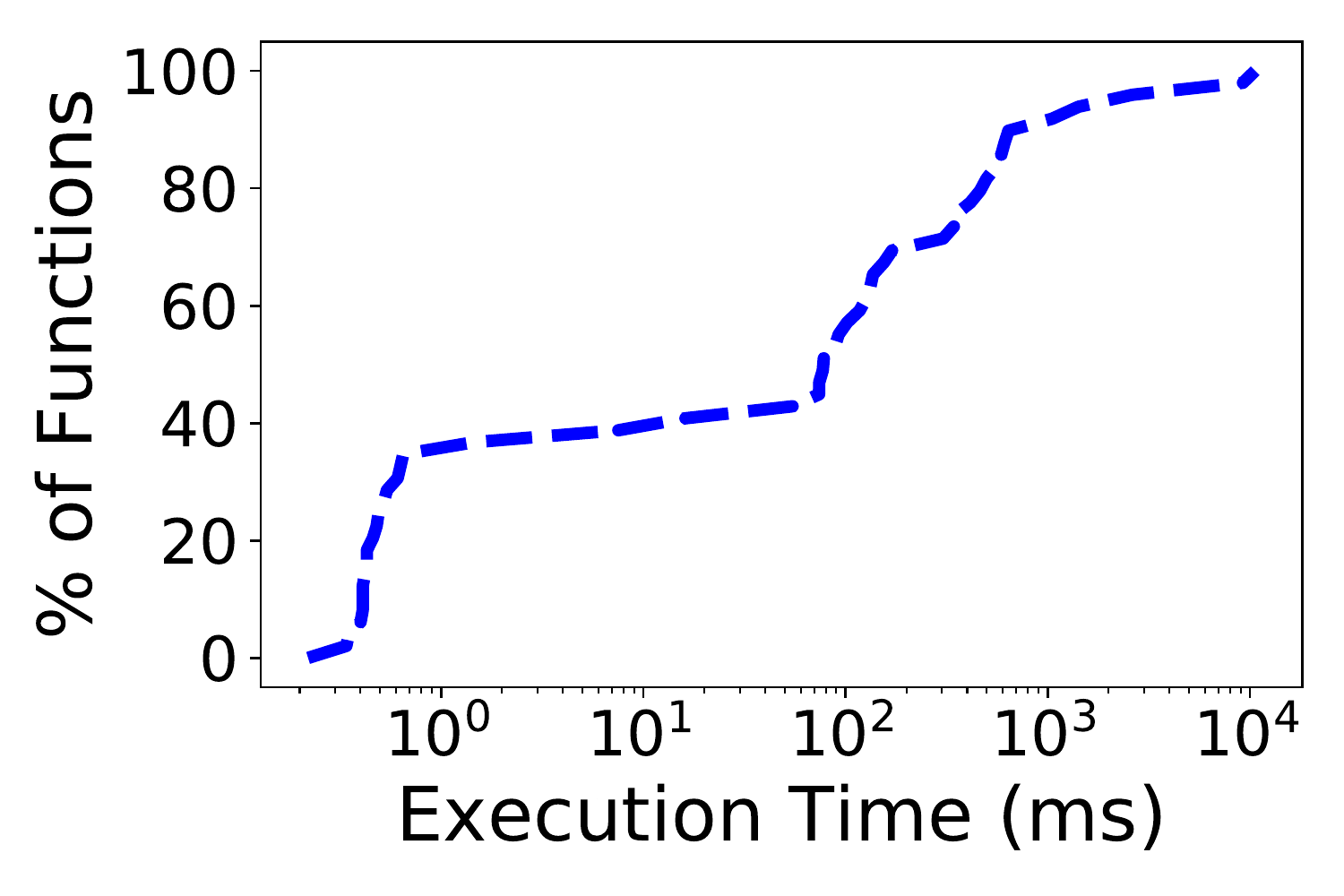}
 		\label{fig:exec_time_cdf}
 	}
 	\subfloat[][]{%
 		\includegraphics[width=0.25\textwidth]{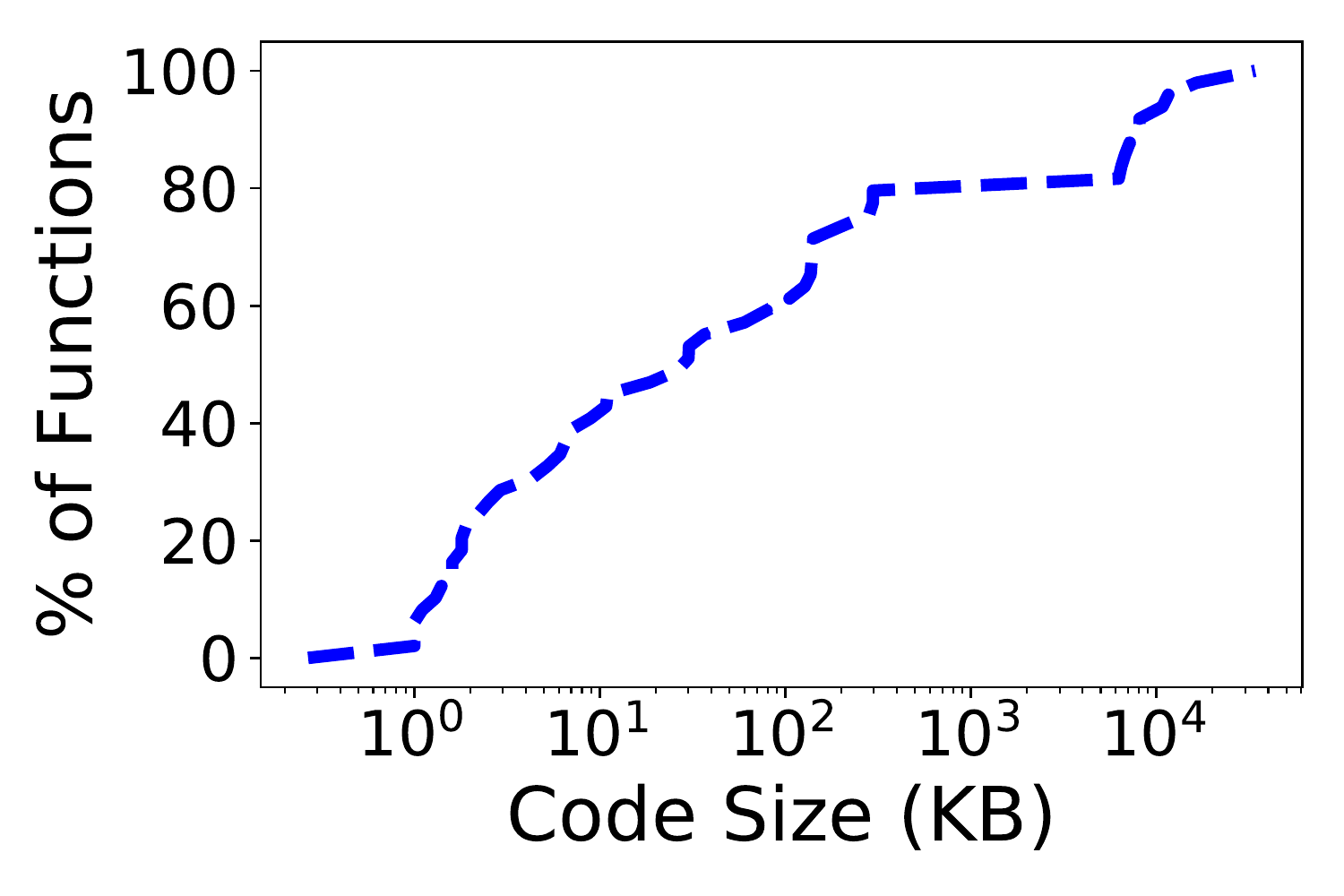}%
 		\label{fig:code_size_cdf}
 	}
 	\subfloat[][]{%
 		\includegraphics[width=0.25\textwidth]{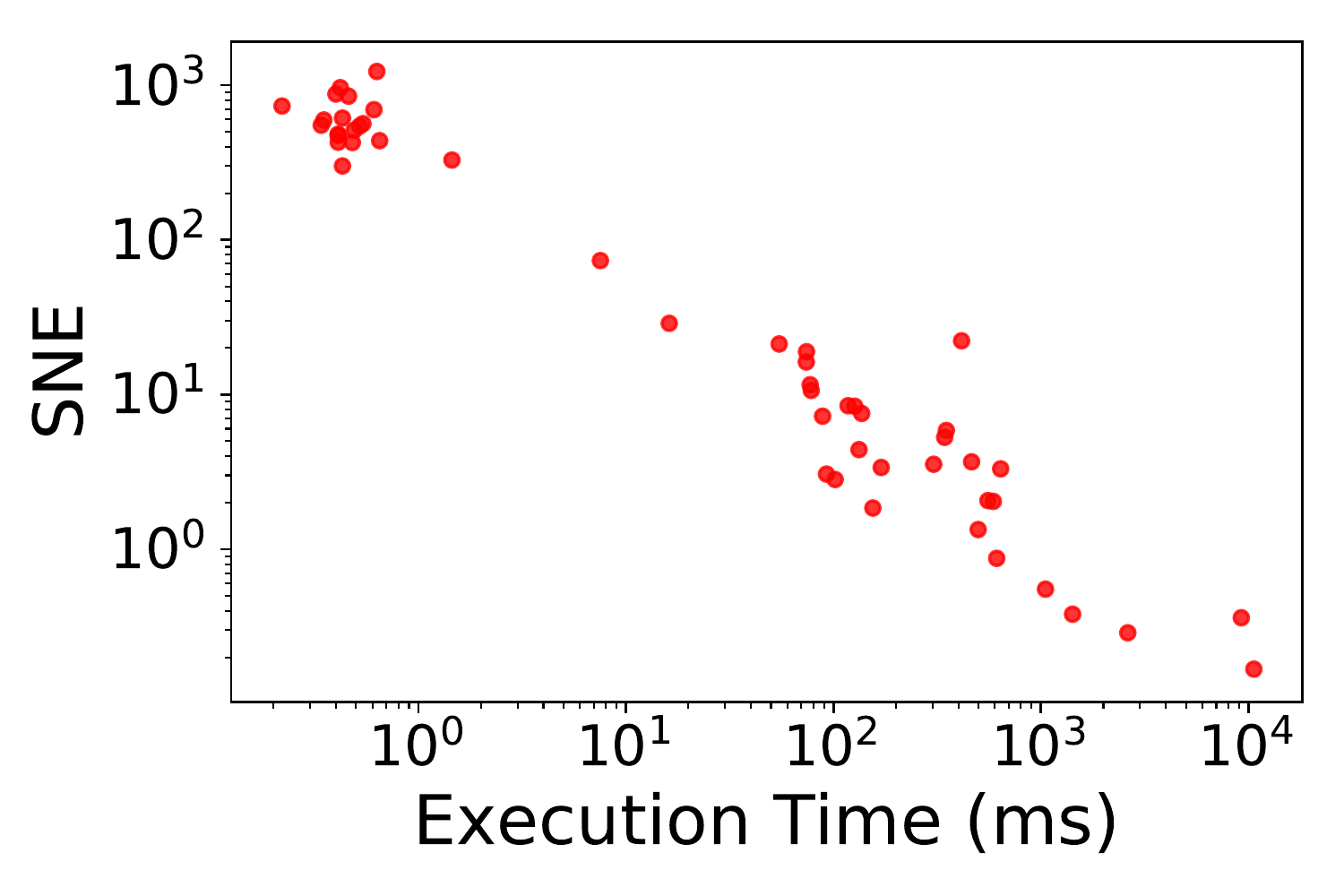}%
 		\label{fig:sne_cdf}
 	}
 	\subfloat[][]{%
 		\includegraphics[width=0.25\textwidth]{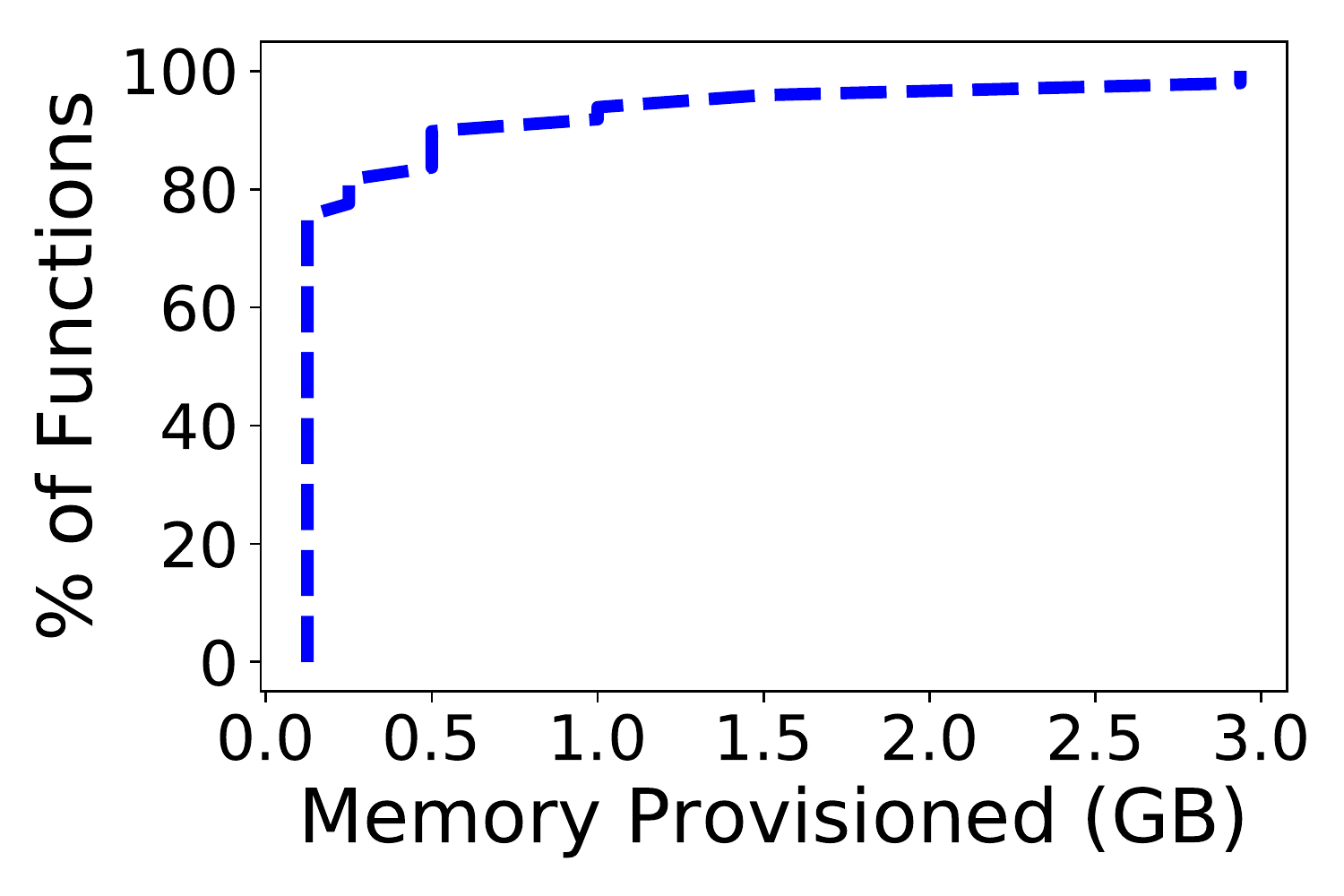}%
 		\label{fig:mem_provisioned_cdf}
 	} 	
  	\vspace*{-2mm}
 	\caption{ \footnotesize Distribution of (a) execution time, (b) code size, (c) SNE and (d) memory provisioned across the 50 functions}
 	\vspace*{-6mm}
 \end{figure*}

  \begin{figure*}[t]
 	 \captionsetup[subfloat]{captionskip=-3pt}
 	\centering 
 	\subfloat[][]{%
 		\includegraphics[width=0.25\textwidth]{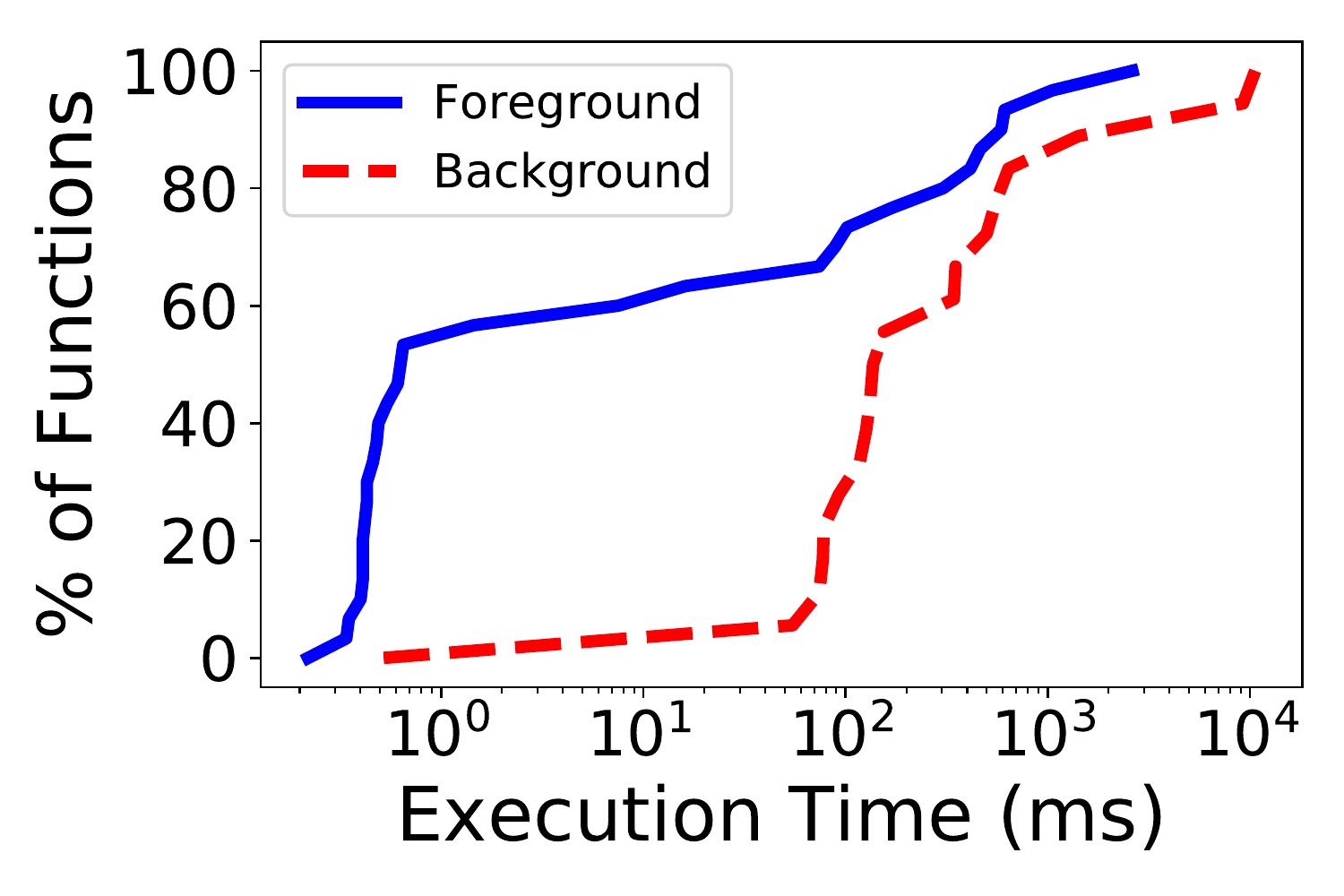}%
 		\label{fig:exec_time_job_types_cdf}
 	}
 	\subfloat[][]{%
 		\includegraphics[width=0.25\textwidth]{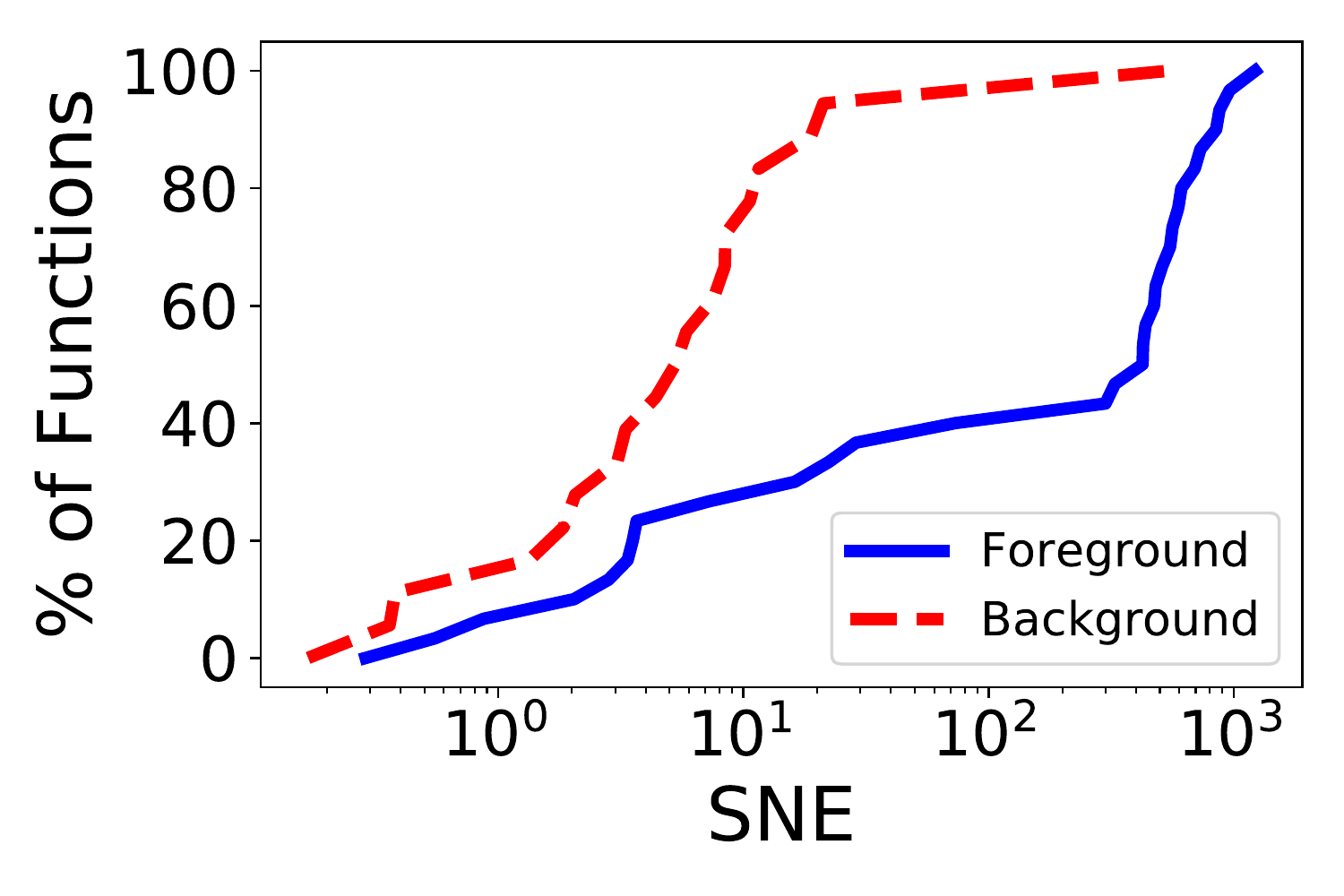}%
 		\label{fig:sne_job_types_cdf}
 	}
 	\subfloat[][]{%
 		\includegraphics[width=0.25\textwidth]{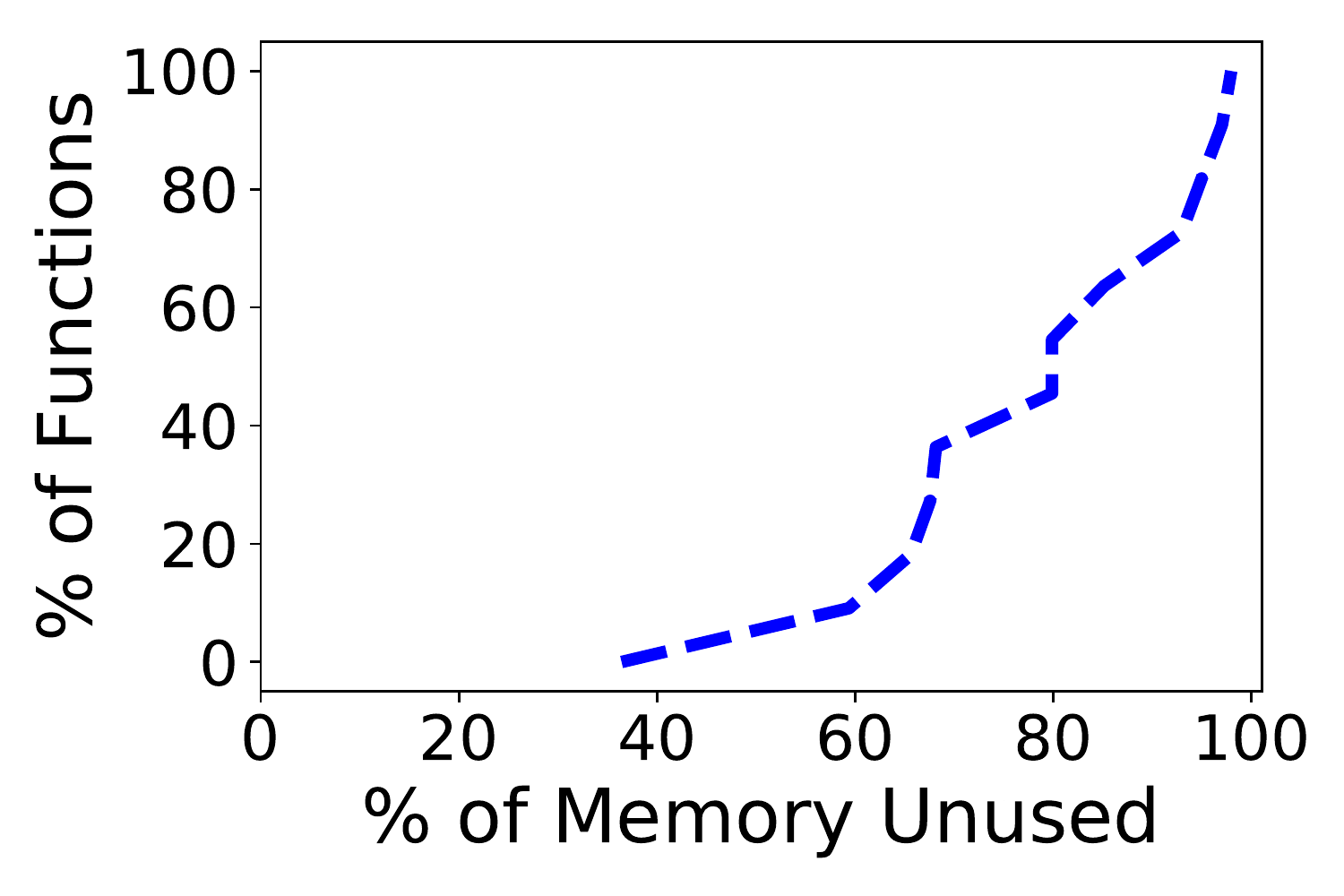}%
 		\label{fig:mem_unused_cdf}
 	} 
 	\subfloat[][]{	
		\setlength\tabcolsep{4pt}
		\begin{tabular}[b]{ c c c } 
			\hline
			&  \footnotesize\textbf{50\%ile} & \footnotesize\textbf{99\%ile} \\
			\hline
			\footnotesize FIFO & \footnotesize 100ms & \footnotesize 101ms \\ 
			\hline
			\footnotesize Sparrow & \footnotesize 102ms & \footnotesize 136ms \\
			\hline
         &   \\
         &   \\
         &   \\	
         \label{table:sparrow}		
		\end{tabular}

    }
  	\vspace*{-2mm}
 	\caption{ \footnotesize Distribution of (a) execution time, (b) SNE across the foreground and background functions and (c) memory unused across functions that provisioned greater than 128 MB. (d) end-to-end latency comparison between FIFO and Sparrow when the incoming workload leads to a cluster CPU utilization of \textasciitilde70\%.}
 	\vspace*{-4mm}
 \end{figure*}

We characterize serverless workloads by studying the top 50 deployed apps (as of November 1, 2019) in
the AWS Serverless Application Repository (SAR)~\cite{sar}.

SAR consists of diverse apps that run on AWS Lambda~\cite{awslambda}. Internally,
AWS Lambda uses Firecracker microVMs~\cite{firecracker} to run the apps. These apps 
typically interact with other AWS services (e.g., S3~\cite{s3})
as well as third-party services (e.g., Slack~\cite{slack}). 
This repository is widely used by the serverless
community which is evident from the fact that the top app has been deployed 45K
times. All 50 apps have a single function, but many recent serverless proposals have
rich DAGs of functions~\cite{excamera,pywren,numpywren,sprocket}. Out of the 50 functions studied, 23 are in NodeJS, 26 in Python, and 1 in Java. 

\noindent\textbf{Benchmarking Methodology.} We use the AWS CLI to upload and trigger the
execution of the functions under study. The functions were triggered to run in
the us-east-1 region via a VM running in the same region. We collect the
following statistics:
(1) function {\em code size}; (2) {\em provisioned memory} - memory 
available to the function during execution as configured by the programmer while
uploading the function to the platform ; (3) {\em runtime memory} - actual memory
consumed during function execution; (4) {\em sandbox setup overhead} - time
taken to setup the function sandbox which includes the steps discussed above;
and (5) {\em execution time} - time taken to execute the core function logic
(without including the sandbox setup overhead). Finally, we also classify functions
as {\em foreground} (typically serving user-facing apps) or
{\em background} based on what they are intended for.

We next discuss the key takeaways from our analysis:

\noindent\textbf{[T1] Functions have a wide range of execution times.}
As seen in Figure~\ref{fig:exec_time_cdf}, 57\% of functions have an
execution time of less than 100ms. These typically corresponds to
user-facing functions (e.g.,
\textsc{alexa-skills-kit-nodejs-factskill}). Also, ~\textasciitilde
10\% functions have an execution time $>1$ second (e.g.,
\textsc{NYC-Parks-Events-Crawler} takes
~\textasciitilde 10s). Additionally, recent works in academia have
shown that serverless platforms are attractive for embarrassingly
parallel tasks that can last for even longer
durations~\cite{pywren,numpywren,excamera} (~\textasciitilde 100s). 
Fig.\ref{fig:exec_time_job_types_cdf} further shows the split of execution
times based on whether they are foreground and background. As
expected, we see that majority (\textasciitilde 65\%) of the
foreground functions have execution times $<100$ms whereas
background functions typically run longer with fewer than
~\textasciitilde 5\% having execution times $< 100$ms.

\noindent\textbf{[T2] Functions have a wide range of code sizes.} Allocating sandboxes also involves downloading the code from the datastore and 
setting up the runtime. Prior works have shown that these steps can take 
significant amount of time (upto 10s of seconds) depending on the code~\cite{sock}. In our analysis
(Fig.~\ref{fig:code_size_cdf}), we notice that code sizes can be as large as
34MB. 

\noindent\textbf{[T3] Sandbox setup overheads dominate execution times.} We
measure the ratio of the sandbox setup overhead to the execution time of the
apps to investigate the impact of overheads on the end-to-end latencies. We
refer to this ratio as {\em SNE} (sandbox setup overhead
normalized by execution time). Fig.~\ref{fig:sne_cdf} indicates that sandbox
setup overheads dominate for $>$ 88\% of the functions with the overhead
being $>$ 100$X$ in 37\% of them. Our observations are consistent
with data from prior work \cite{sock,pipsqueak,serverless-peek}.
Fig.~\ref{fig:sne_job_types_cdf} shows that high sandbox setup overheads
impact foreground functions much more severely.

\noindent\textbf{[T4] Functions typically have small memory footprints.}
Fig.~\ref{fig:mem_provisioned_cdf} shows the maximum memory provisioned by the
functions. 78\% of the them require only 128MB. Fig.~\ref{fig:mem_unused_cdf}
further shows that most functions requesting more than 128MB of provisioned
memory typically leave a significant fraction of provisioned memory unused.

\noindent\textbf{[T5] Majority of apps have a single function.} All of the top 50 deployed apps have only a single function. Out
all the apps on SAR, we found only two instances of DAGs
which were a linear chain of 2 functions (e.g.,\textsc{cw-logs-to-slack}). However, as noted earlier, many emerging applications induce richer DAGs. Our work aims for generality, and thus our work also encompasses applications that are DAG-structured, as opposed to focusing on single function ones.

\subsection{Serverless Platform Requirements} 
Based on the above takeaways, the requirements of an ideal serverless platforms are as follows:

\noindent\textbf{[R1] Minimize the impact of sandbox setup overheads on end-to-end request latencies:} Given that these  overheads dominate 
execution times (T3), we wish to eliminate them from end-to-end request execution critical paths. 
  
\noindent\textbf{[R2] Minimize the impact of control plane overheads on end-to-end request latencies:} Given that functions with low execution times are the
common case (T1), we require the load balancing and scheduling layers of
the platform to make decisions in sub-millisecond at scale.

\noindent\textbf{[R3] Have a scalable control plane:} Given that many apps will 
use the platform and their request load can grow high arbitrarily, we require scalable load balancing and scheduling where neither can become a bottleneck.

\noindent\textbf{Overall Goal.} Given that many applications may run
simultaneously on the platform, our high-level goal is to support
tight performance bounds for application requests. Specifically, we
wish to ensure that, per application, end-to-end
latencies are ``close'' to native application execution times for a
vast majority 
of requests
We allow developers to define how
``close'' to native execution they wish to be, by allowing them to
specify a deadline. 
\subsection{Issues with Serverless Platforms Today}

Existing platforms and mechanisms cannot meet the above goal due to:\\ 
\noindent\textbf{1. Reactive, Fixed, and Workload-Unaware Sandbox Management Policy.} Most of
today's serverless platforms~\cite{openwhisk,awslambda,AzureFaas,GCF} only reactively setup sandboxes, i.e., the scheduler waits for a
request to arrive and only then sets up a sandbox (if existing ones are busy)
leading to requests experiencing additional latency. Also, given the
overheads associated, to amortize the overheads across future requests, platforms adopt a static and workload-unaware policy - a
sandbox is kept loaded in memory for a fixed amount of time (since its last invocation).
While the above policy is simple to implement, it
does not work well in practice as - (a) it does take into account workload
characteristics while making decisions which can lead to wasteful memory consumption (e.g., when sandboxes are loaded even when the workload does not require
them), or additional overheads (e.g, too few sandboxes available and workload
increases suddenly); and (b) is easy to game for external users (e.g., frequently send
dummy requests to ensure that the sandbox is not evicted~\cite{serverless-game}). 

\noindent\textbf{2. Sub-Optimal Scheduler Architectures.} While
centralized schedulers can make optimal scheduling
decisions, when incoming workload grows arbitrarily, a centralized
approach can easily become a scalability bottleneck. Decentralized
approaches are promising, but they trade-off scheduling quality or low
predictable scheduling latencies for achieving scalability, which lead
to higher end-to-end latencies.

For instance, {\em parallel global scheduling}
approaches (e.g., Sparrow~\cite{sparrow}), where multiple schedulers with a
global view carry out scheduling by randomly probing two machines, may not
find the best-fit for the function under load as it randomly probes machines
 and does not make an optimal scheduling decision (Fig.~\ref{table:sparrow}). Similarly, {\em bottom-up
hierarchical scheduling} (e.g., Ray~\cite{ray}), where functions are
first submitted to a per-node local scheduler and are sent to a randomly
chosen global scheduler only when it is not possible to schedule locally (say
due to overload), may experience unpredictable scheduling latencies as the
function may bounce back and forth between node and global schedulers due to
conflicts between multiple global schedulers.
 
\noindent\textbf{3. Homogeneous Request Handling.} In serverless platforms today, every incoming request is handled in the same manner, which limits them from making intelligent scheduling decisions. In practice, functions have varying latency requirements; e.g., foreground functions are typically latency sensitive and can tolerate limited additional delay, whereas background functions normally have higher slack and can tolerate higher delay. And, not all functions with tight latency requirements are likely to impose high load at the same time. An ideal platform can leverage these aspects to carefully multiplex and schedule requests to maximize the number of requests that get their responses before their available slack runs out.  
\section{Key Ideas and Architecture}
\label{sec:architecture}

We now describe the key ideas that form the basis of \name{}, a
serverless platform designed to meet specified deadlines for
latency-sensitive DAG-structured serverless applications running on a fixed-size cluster.

\noindent \textbf{1. Decoupling sandbox allocation from request
  scheduling:} \name{} removes sandbox allocation overhead
(\secref{sec:motivation}) from the critical path of request execution by {\em proactively allocating sandboxes}
ahead of time based on the expected future load for a function.
Additionally, \name{} uses a novel {\em even placement} approach to spread sandboxes across the
cluster so as to maximize the probability of future requests
benefitting from these provisioned sandboxes (\secref{ss:alloc}).
 
\noindent \textbf{2. Autonomous schedulers and SLA aware scheduling:} To scale scheduling, we introduce {\em semi-global schedulers} (SGSs).
Each SGS is responsible for exclusively managing a partition
of the cluster machines known as its {\em worker pool}. This ensures that a scheduler does not become a scalability bottleneck and ensures that schedulers make
optimal decisions within the worker pool.
We also develop a {\em deadline-aware} scheduling strategy (\secref{ss:sched})
that leverages the flexibility of the different slack requirements
amongst requests and multiplexing among apps' requests
(\secref{sec:motivation}) to ensure that deadlines are met.

\noindent \textbf{3. Co-designing the load balancing and scheduling layers:}
Partitioning the cluster into a number of SGSs introduces 
the challenge of determining which DAGs are assigned to which SGS. We use the load balancer to address this challenge and 
codesign the load balancing and scheduling layers so that the load balancing
layer has the required visibility to (a) do {\em sandbox-aware} request routing and
(b) prevent individual SGSs from becoming hotspots. Doing
so maximizes future requests that benefit from proactive allocation.
Additionally, we develop a low-overhead {\em gradual scaling} mechanism 
that allows logically scaling up/down the schedulers associated with a DAG to prevent hotspots (\secref{ss:scaling}) without unduly impacting request processing.

We next present an end-to-end example that highlights the various features of \name{}. 

\noindent \textbf{Initial DAG Upload.} The user develops the functions that
make up the computation DAG and uploads them to our platform. During the initial upload,
as done today, the user also specifies the resource requirements of the
functions along with the DAG structure using a JSON-based language. Crucially,
we also require the user to specify the maximum execution time for 
the DAG given a new input trigger. This can be derived from the 99\% percentile latency 
that is acceptable for an application. \name{} aims to maximize the number of requests
that are completed within this deadline.

\begin{figure}[t]
    \centering \includegraphics[width=0.85\columnwidth]{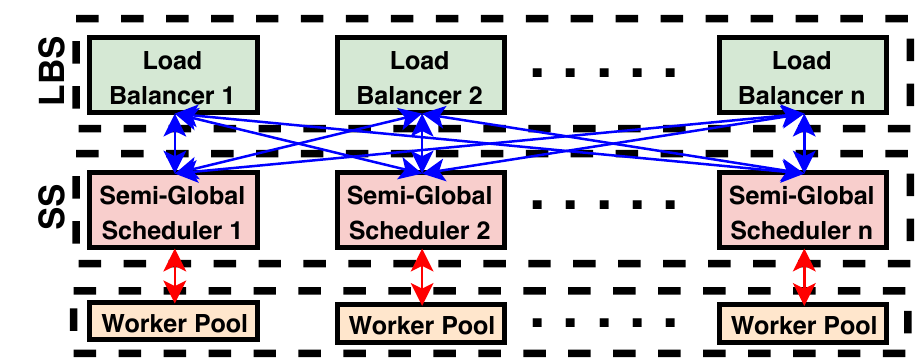}
    \caption{
        \name{} Architecture. Core services include load balancing service and a scheduling service consisting of semi-global schedulers that manage their own worker pool.
    }
    \label{fig:overview}
    \vspace{-0.1cm}
\end{figure}

\noindent \textbf{Request Control Flow (see Fig.~\ref{fig:overview}).} When a 
request arrives at our platform, it gets routed to one of the many load balancers (LB) 
that form the load balancing service (LBS). The LB routes it to one of the many SGSs that form the scheduling service (SS) based on its routing
policy. At the SGS, the request is enqueued for scheduling. Requests are prioritized
by the SGS in a deadline-aware fashion
and run on available workers in the worker pool in a work-conserving fashion.

In the background we perform two main actions: first, the SS monitors the memory available 
and the incoming traffic to adjust the sandbox allocations and places sandboxes so as to maximize the benefit of proactive allocation. Second, the LBS monitors the
load on each SGS and adjusts the routing policy accordingly.
We discuss the details of each of the above mentioned components in subsequent sections.

\section{Scheduling Service (SS)}
\label{sec:scheduling-service}

SS is responsible for managing sandboxes and
scheduling incoming DAG requests. We first describe the architecture
that makes it scalable (\secref{ss:sgs-arch}) and then discuss the deadline-aware strategy used to minimize deadlines missed 
(\secref{ss:sched}). Finally we explain the
approach used to proactively allocate sandboxes so as to minimize the
impact of sandbox setup overheads(\secref{ss:alloc}).

\subsection{Semi-Global Schedulers (SGS)}
\label{ss:sgs-arch} 

To handle the low latency requirements and make optimal scheduling decisions,
\name{} divides the cluster into a number of \emph{worker pools}, where each worker pool
consists of a subset of machines in the cluster. Every worker pool is then assigned to a semi-global scheduler (SGS) and these
semi-global schedulers form a part of the scheduling service.

Given the nature of our workload, where we have a small number of independent, latency-sensitive DAGs
, we partition the DAGs such that each SGS is only responsible for
a subset of DAGs. This assignment can change at a coarse-time granularity and is managed by the load
balancing service.

\noindent\textbf{Sizing Worker Pools.} While deploying \name{}, 
the platform admin is responsible for determining the size of each
worker pool. The trade-off here is that using too large of a worker pool would lead
to increased scheduling delays (as discussed in \secref{sec:motivation}). On the other hand
using too small a worker pool could result in load imbalance across various SGS and necessitate
frequent load balancing (\secref{subsec:sens_analysis}). As an extreme, if we choose a worker pool with just a single
machine then the load balancer would need to perform all the scheduling of requests.
A simple approach we espouse is to organize each rack as a worker pool 
with one of the machines running the SGS. 

\subsection{Deadline Aware Scheduling} 
\label{ss:sched}
We next present the strategy we use to schedule requests in an SGS, first in the context of 
individual functions and the generalize it to requests traversing a DAG.
Requests are routed to an SGS from the load balancer and incoming requests are placed in a 
scheduling queue. Given our goal of meeting latency deadlines, we would like to adopt a 
scheduling policy that minimizes the number of missed deadlines. Additionally, given the short execution times, we assume that functions cannot be pre-emptied during execution.

Following classic scheduling approaches to minimize the execution
time~\cite{HarcholBalter2003SizebasedST,schrage1968}, 
we propose using the \textit{shortest remaining slack first (SRSF)} algorithm. Whenever
a CPU core becomes available, the SGS filters requests to only consider ones whose 
resource requirements are met by the current available resources and then calculates 
a \emph{remaining slack} for the filtered requests.
Slack here is defined as the
time a function request can be queued without violating its deadline.

The SGS
prioritizes and picks the function request that has the least remaining slack.
In case of ties, the SGS picks the function which has the least remaining work.
Doing so ensures that we quickly get another opportunity to schedule, which
further minimizes deadlines missed. Additionally, scheduling based
on remaining slack also avoids starvation for requests with large amount of slack.
Finally, the SGS schedules requests on available workers in
a work-conserving manner. The SGS spreads out sandboxes for a function across its workers to maximize the chances that a proactively allocated sandbox will be available at the worker (~\secref{sss:placement}).

\noindent\textbf{DAG Awareness.} We now extend the scheduling strategy
to handle a DAG. Given the user-specified DAG deadline, the key question
that needs to be answered is, how is the remaining slack calculated for a 
DAG?
After a function is processed, the remaining slack for each function of a DAG is calculated by subtracting the critical path execution time~\cite{cpm,critical-path-def} from the time remaining to the DAG's deadline.
As an SGS is DAG aware,
it schedules functions once their dependencies are met by calculating the
RS in the manner stated.

\subsection{Proactive Sandbox Allocation}
\label{ss:alloc}

\begin{figure*}[t]
	\captionsetup[subfloat]{captionskip=0.0pt}
	\centering 
	\subfloat[][]{%
		\hspace{-1cm}
		\includegraphics[scale=0.37]{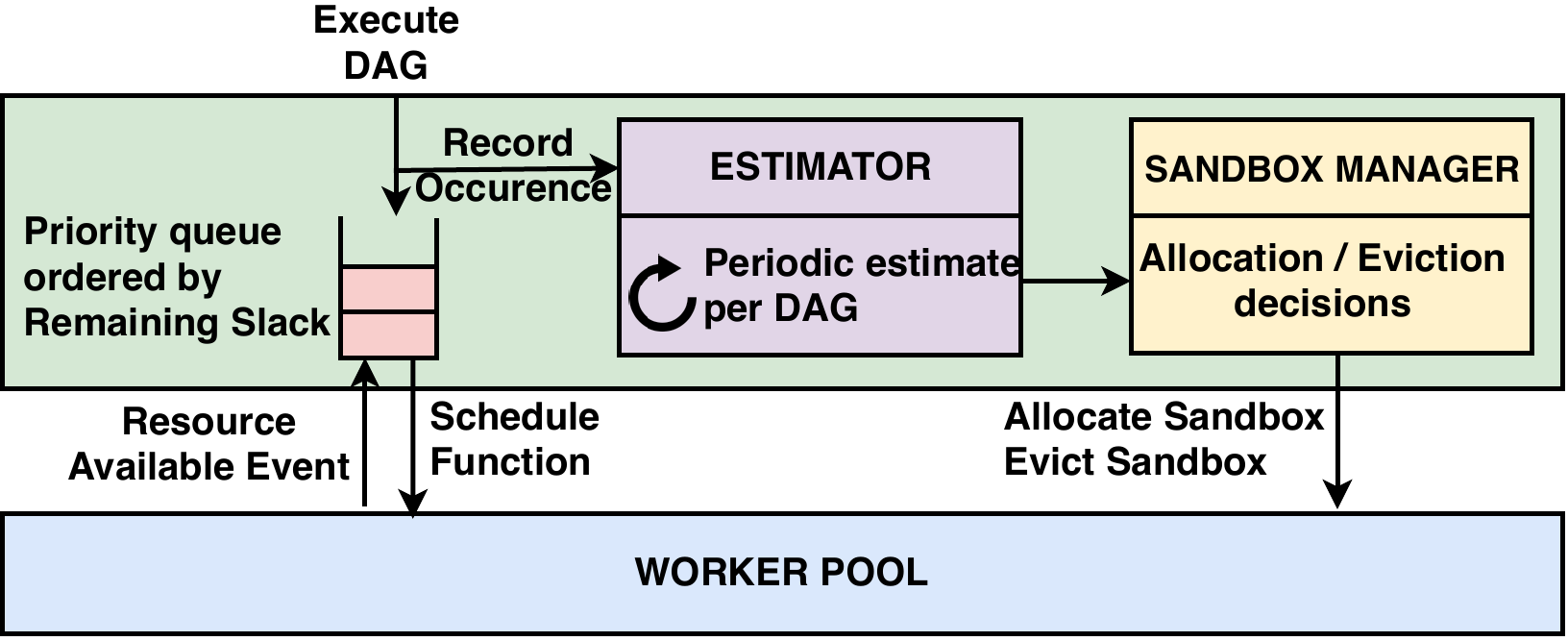}%
		\label{fig:sgs-arch}
	}
	\hspace{0.3cm}
	\subfloat[][]{%
		\includegraphics[scale=0.40]{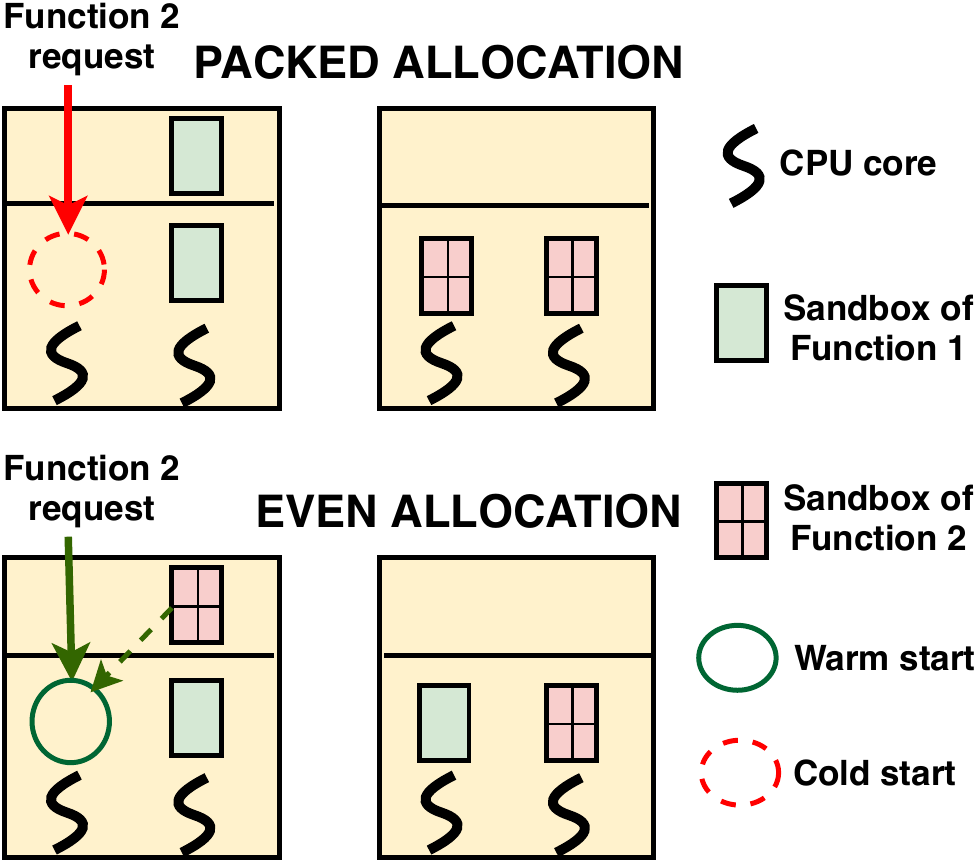}
		\label{fig:alloc}
	}
	\hspace{0.3cm}
	\subfloat[][]{%
		\includegraphics[scale=0.35]{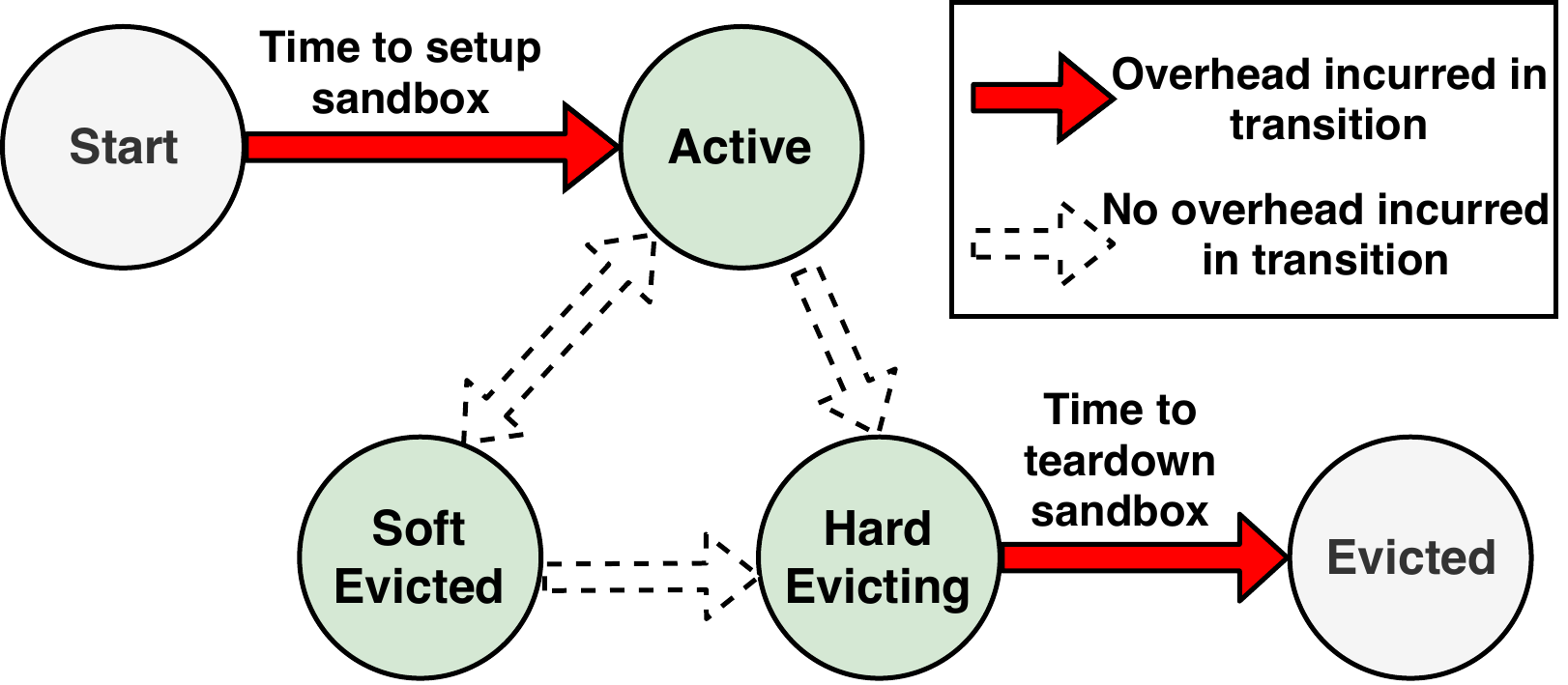}%
		\label{fig:lifecycle}
	}
	\vspace*{-1mm}
	\caption{ \footnotesize (a) Zoomed-in view of a semi-global scheduler consisting of a priority queue, an estimator, and a sandbox manager (b) Comparison of multiplexing in packed and even allocation policies. With packed allocation, the execution of function 2 incurs a cold start (marked in dashes) due to the unavailability of a proactively allocated sandbox on that machine. With even allocation, the execution of function 2 does not incur a cold start (marked in solid) since a proactively allocated sandbox is available (c) State diagram showing transitions between different stages of the sandbox lifecycle along with overheads incurred}
	\vspace*{-4mm}
\end{figure*}

Given that typical serverless workloads have their execution time in the same
order of magnitude as that of setting up sandboxes (\secref{subsec:real_world_study}), we need to ensure
that requests are not exposed to this overhead. To
achieve this, \name{} decouples sandbox allocation from scheduling of incoming
requests and this allows each SGS to proactively setup sandboxes across its worker pool
based on the future expected load. This is in contrast to today's
platforms~\cite{openwhisk} that are not workload-aware and reactively setup
sandboxes when a request arrives. By decoupling sandbox allocation from scheduling, \name{} promotes the pipelining of sandbox allocation with scheduling decisions resulting in reduced impacts of cold starts.

Proactively allocated sandboxes occupy memory and do not consume any other
resources. With high-memory machines becoming the norm and serverless
functions having small memory footprint (\secref{sec:motivation}), we believe
it is viable to trade off the memory consumed by the proactively allocated
sandboxes to ensure that users are not exposed to sandbox setup overheads.
To limit the amount of memory used, 
the platform administrator can configure the amount
of memory on each machine that can be used to proactively setup sandboxes. We
refer to this memory as the {\em proactive memory pool} from here on.
Finally, we note that proactively allocated sandboxes are a form
of \emph{soft state}~\cite{fox1997cluster} that can potentially improve performance without affecting 
correctness.

Each SGS is responsible for proactively setting up sandboxes of functions for
which it is receiving requests (as decided by the LBS). In order to do so, the
SGS must answer the following questions: (1) how many sandboxes of each
function must be setup proactively? (2) how should these sandboxes be placed
on its worker pool? (3) when/how should these sandboxes be evicted from the
proactive memory pool?

\subsubsection{Sandbox Demand Estimation}
\label{sss:sandbox_demand}

\begin{figure}[t]
	\centering \includegraphics[scale=0.5]{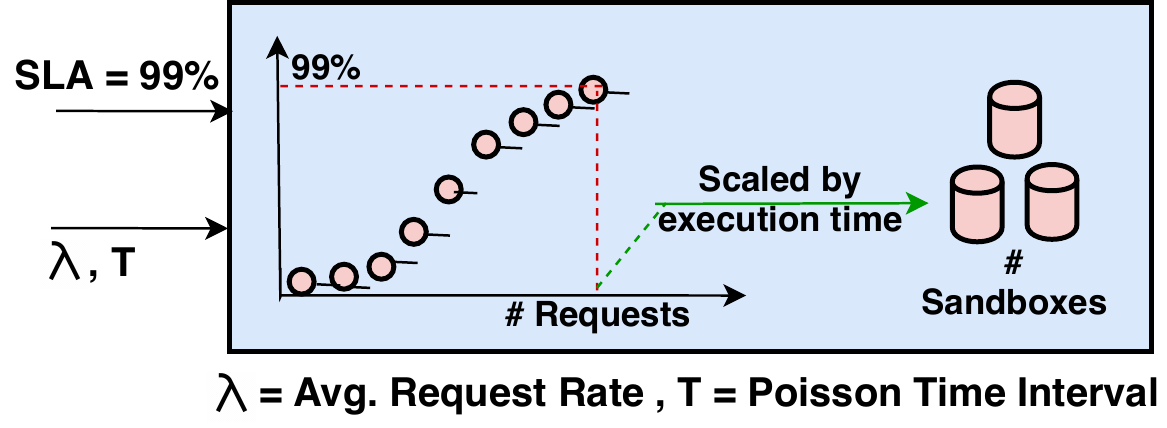}
	\caption{
		Estimating number of sandboxes to proactively allocate
	}
	\label{fig:poisson}
	\vspace{-0.1cm}
\end{figure}

For each DAG that is being handled by the SGS, our goal is to determine
the minimum number of sandboxes that need to be allocated for each of its
constituent functions, so as to meet the agreed upon SLA. Given the execution
time of a function and the SLA, we model how requests of the function arrive to determine the minimum number of sandboxes needed.

In \name{}, we make an assumption that request inter-arrival times
follow an exponential distribution and model the number of requests expected
in a given time interval $T$ as a Poisson distribution. Specifically, given the
SLA (e.g., 99\%), we use the inverse distribution function to find the maximum number of
requests that can arrive in T (Fig.~\ref{fig:poisson}).
However, given that execution time of a function can be longer
than T, we scale up the maximum number of requests to account for requests
that overflow from the current time interval to the next one.

The SGS requires an estimate of the arrival rate of a function, so as to
construct the Poisson distribution, which can then be used to determine the
number of sandboxes using the above approach. In the background, the SGS (via its estimator
module, Fig.~\ref{fig:sgs-arch}) continuously records the arrival rate of the function (over a 100 ms
interval in our prototype) and uses an exponentially weighted moving average
(EWMA) over the current interval's measured rate and the previous estimate to
get the new estimate. The SGS measures and estimates this for all the functions
that it is handling.

\subsubsection{Sandbox Placement}\
\label{sss:placement}
Now, given the number of sandboxes that need to be setup proactively for a
function, the SGS needs to decide how to place these sandboxes across the
various workers in its worker pool. Ideally, we would want to place the
sandboxes to maximize the number of future requests that will use them.

Given recent efforts~\cite{sock} towards reducing the memory footprint of proactively
setting up sandboxes, a tempting approach would be to pack as many sandboxes
of the same function on the same worker. While this reduces the memory
overhead, it does not increase the probability of future requests benefiting
from proactive allocation. For example, consider a scenario where there
are two worker machines and the demand estimation of two
functions is 2 sandboxes each. Using the above approach, the sandboxes
belonging to the same function are setup on the same worker (see
Fig.~\ref{fig:alloc}). In such a case, when a core becomes available on
worker one and the outstanding request for the second function is to be
scheduled, it experiences the overhead of setting up a new sandbox as no
compatible sandbox is available on the worker.

Instead, in \name{}, for a given function, we {\em evenly} spread its sandboxes
across the various workers (lines 18-38 in Pseudocode~\ref{alg:sandbox-mgmt}).
Specifically, given the number of sandboxes required, for each sandbox that
needs to be setup, the following 2-step process is taken (via the allocator
sub-module, Fig.~\ref{fig:sgs-arch}): (1) determine the worker that has the
minimum number of sandboxes of this function, and (2) setup sandbox on the
worker. This approach 
improves statistical multiplexing, i.e., makes it easier for
future requests to find a proactive sandbox. In Fig.~\ref{fig:alloc}, the request does not incur setting up overhead as a compatible
sandbox is available.

\floatname{algorithm}{Pseudocode}
\begin{algorithm}[t!]
\begin{scriptsize}
\begin{algorithmic}[1]

\State{$\triangleright$ Given a DAG D, either allocate or evict sandboxes}
\Procedure{SandboxManagement}{DAG D}
	\State $\mathbb{M}$ \Comment{Mapping between DAG and demand}
	\State oldDemand = $\mathbb{M}$[D.id]
	\State newDemand = D.demand
	\If{newDemand $>$ oldDemand}
		\State{$\triangleright$ Allocate sandboxes as demand increased}
		\ForAll{\textsc{f} $\in D.functions$}
			\State \textsc{AllocateSandboxes(f, newDemand - oldDemand)}
		\EndFor
	\ElsIf{newDemand $<$ oldDemand}
		\State{$\triangleright$ Soft evict sandboxes as demand decreased}
		\ForAll{\textsc{f} $\in D.functions$}
			\State \textsc{SoftEvictSandboxes(f, oldDemand - newDemand)}
		\EndFor
	\EndIf
\EndProcedure

\Statex

\State{$\triangleright$ Given a function F and its demand, allocate sandboxes}
\Procedure{AllocateSandboxes}{Function F, Int allocDemand}
	\For{$\_$ \textbf{in} range(allocDemand)}
		\State{$\triangleright$ Get worker which has min sandboxes for this function}
		\State minW = \textsc{getWorkerWithMinSandboxes(F.id)}
		\State{sandboxFound, sandbox = minW.getSoftEvictedContainer(F.id)}
		\If{sandboxFound}
			\State{$\triangleright$ Preferentially allocate a soft evicted sandbox} 
			\State minW.SoftAllocate(sandbox)
			\State \textbf{continue}
		\EndIf
		\If{minW.hasEnoughPoolMem(F)}
			\State{$\triangleright$ Allocate a new sandbox if enough memory available} 
			\State minW.Allocate(F)
		\Else
			\State{$\triangleright$ Otherwise evict a sandbox and allocate}
			\State minW.HardEvict(F)
			\State minW.Allocate(F)
		\EndIf	
	\EndFor
\EndProcedure

\Statex
\State{$\triangleright$ Given function F, evict enough sandboxes to launch a sandbox of F}
\Procedure{HardEvict}{Function F}
	\While{w.freePoolMem $<$ F.memNeeded}
		\State victimF = w.getVictimF() \Comment{Get function based on fairness metric}
		\State w.Evict(victimF)
		\State w.freePoolMem += victimF.memNeeded
	\EndWhile
\EndProcedure

\end{algorithmic}
\end{scriptsize}
\caption{\name Sandbox Management}
\label{alg:sandbox-mgmt}
\end{algorithm}

\subsubsection{Sandbox Eviction}
\label{sss:eviction}
The previous section described how an SGS proactively allocates containers based on estimations. However, when
the estimations deem that not all the sandboxes previously allocated are
required, we need to decide what should be done with these excess sandboxes. A
natural approach would be to evict these containers from the underlying worker pool as
they consume memory. However, in \name{} we 
{\em lazily} evict containers from the worker pool to avoid
unnecessary sandbox allocation overheads.

In \name{}, a sandbox goes through two stages of eviction - \textit{soft
eviction} and \textit{hard eviction} (Fig.~\ref{fig:lifecycle}). When the estimates fall below what was
previously estimated, the SGS marks the excess sandboxes as soft evicted,
i.e., they will not be considered while scheduling requests. Given the excess
number of sandboxes of a function that need to be soft evicted, the SGS needs
to decide which sandboxes across the various workers need to be soft evicted.
For this, the SGS follows a process similar to the placement approach
it takes, with the only difference being that it selects the worker(s) that
have the maximum sandboxes of this type, and {\em soft evicts} a sandbox from it.
This process is repeated until the required number of sandboxes are soft
evicted(lines 11-15 in Pseudocode~\ref{alg:sandbox-mgmt}). The aforementioned approach balances the sandboxes across workers to
the extent possible which improves statistical multiplexing
Having soft evicted sandboxes enables \name{} to deal with
temporary load fluctuations in a better manner.
In such scenarios, sandboxes are soft evicted when the load decreases. When the load increases back, soft evicted containers just need to be unmarked and this incurs no overheads.

Finally, a sandbox is {\em hard evicted} only when the proactive memory pool on a
worker is saturated and a new sandbox needs to be proactively allocated (lines 39-46 in Pseudocode~\ref{alg:sandbox-mgmt}). The
SGS hard evicts the sandbox of a function whose current allocation is closest 
to its estimation. This prevents functions whose allocations are far from their estimation 
being negatively impacted. Also, the SGS prefers to hard evict a soft
evicted sandbox first before evicting a sandbox that may be reused for
scheduling.

\section{Load Balancing Service (LBS)}
\label{sec:lbs}

\floatname{algorithm}{Pseudocode}
\begin{algorithm}[t!]
\begin{scriptsize}
\begin{algorithmic}[1]

\State{$\triangleright$ Given a DAG D, determine if scaling is required}
\Procedure{Scaling}{DAG D}
	\State $\myvec{N}$ \Comment{per associated SGS sandbox count for DAG D}
	\State $\myvec{qDelay}$ \Comment{per associated SGS observed queuing delay for DAG D}
	\Statex
	\State weightedQDelay = $\sum_i$ $\frac{\myvec{N}_{i} * \myvec{qDelay}_{i}}{\sum_i \myvec{N}_{i}}$
	\State scalingMetric = $\frac{weightedDelay}{D.slack}$
	\If{scalingMetric $>$ ScaleOutThreshold}
		\State \textsc{ScaleOut(D)}
	\ElsIf{scalingMetric $<$ ScaleInThreshold}
		\State \textsc{ScaleIn(D)}
	\EndIf
\EndProcedure

\end{algorithmic}
\end{scriptsize}
\caption{\name Per DAG SGS Scaling}
\label{alg:scale-mgmt}
\end{algorithm}

The LBS is responsible for routing 
requests to the underlying SGSs. We discuss its responsibilities (\secref{ss:resp}) 
and then discuss how our service performs the
tasks at hand (\secref{ss:scaling}).

\subsection{Service Responsibilities}
\label{ss:resp}
The LBS has two key responsibilities : (1) balance load across SGS: given that the
underlying SGSs partitions the cluster, the LBS should 
ensure that the load is spread across the various SGS and a single SGS does
not become a bottleneck; 
 (2) perform {\em sandbox-aware routing}: given 
that the SGSs proactively allocates sandboxes, the LBS should route
requests appropriately with the objective of maximizing the number of requests
that benefit from the proactive allocation.

\subsection{Scaling SGSs used per DAG}
\label{ss:scaling}
Given that the underlying cluster is partitioned and is managed by various
SGSs, a key question that needs to be answered is among how many SGSs should
the incoming requests of a DAG be spread? A possible solution would be to use
all the available SGSs and spread the incoming requests evenly.
This would avoid hotspots but naively applying such an
approach in our context would lead to degraded performance as more requests
would experience the sandbox allocation overhead as each SGS triggers
allocations only when it starts receiving requests.

At the other extreme is the option of routing all requests of the DAG to a
single SGS. While this approach does not suffer from the same limitations,
a single SGS may not have enough capacity to
handle the incoming workload. 
Thus, we choose a middle ground and dynamically associate the right number
of SGSs that are needed to handle a DAG. 
However, to ensure that this dynamic
approach is effective and performant, the following questions need to be
answered - (1) what should be used as the indicator to scale SGSs in and out? (2)
what is our scaling mechanism? and (3) how do we ensure that the request latencies do not
suffer when we scale out/in?

\subsubsection{What is the scaling indicator?} 
There are a number of situations under which the current number of SGSs
associated with a DAG could be too few, requiring scale out. First, 
when the incoming workload of a DAG cannot be handled by the
current SGSs due to resource unavailability. This can happen either due to the
incoming load being too high or due to contention with other DAGs that are
handled by the same SGSs. Second, we also need to scale out when there is
severe pressure on the cumulative proactive memory pool which can lead to
users experiencing sandbox allocation overheads.

Rather than relying on multiple independent metrics to indicate the occurrence of the
above situations, we leverage queuing delay experienced
by requests (of the corresponding DAG) at the SGS as the universal metric. Queuing delay
covers all the situations and is easily observable.
Specifically, each SGS measures the queuing delays per DAG using EWMA (similar
to how it estimates the per DAG RPS) over a window. Having a window ensures
that our system does not react to transient changes in queuing delays.

The SGS piggybacks this measured queuing delay
with each outgoing response to the LBS. The LBS further uses this information to
decide if we need to scale out/in.

\subsubsection{What is the scaling mechanism?}

\begin{figure}[t]
	\centering \includegraphics[scale=0.45]{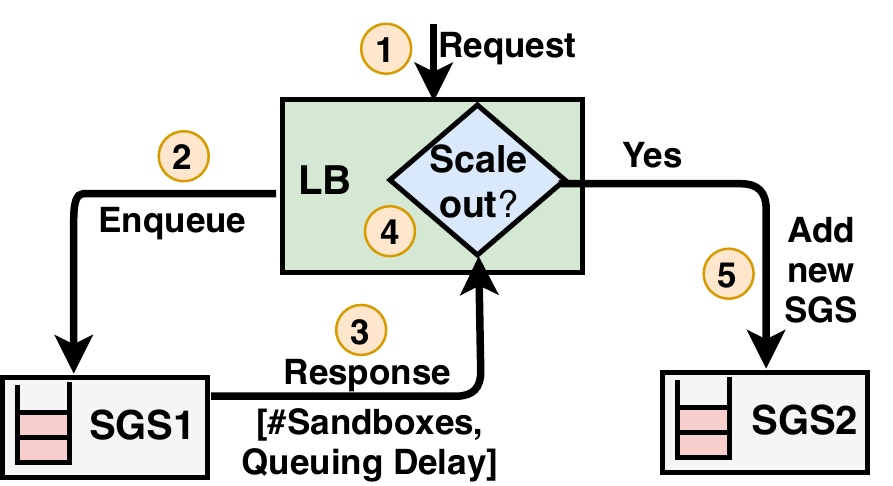}
	\caption{
		Interaction of load balancer with SGSs during a scale out
	}
	\label{fig:lb_scaling}
	\vspace{-0.1cm}
\end{figure}

\noindent\textbf{Initial SGS Selection.} When a request for a particular DAG arrives for the
first time at the LBS, we use consistent hashing~\cite{consistent} to determine which SGS to
route requests to. Specifically, the LBS maintains a consistent hash ring - with
all the underlying SGSs hashed to the ring (by using their ID). Now when the
first request arrives, the LBS hashes the DAG ID to the ring and assigns it
its initial SGS. Using consistent hashing ensures that no single SGS is
overwhelmed by being responsible for a large share of DAGs.

\noindent\textbf{Scale Out (see Fig~\ref{fig:lb_scaling}).} The LBS receives the queuing delay observed by the
requests of this DAG at the various SGSs. It then
computes a scaling metric which is a function of the reported per-SGS
queuing delays normalized by the deadline (described below). If the metric is above a {\em scale-out
threshold}, then the LBS scales out by associating another SGS (the next one in
the ring) with this DAG (lines 7-8 in Pseudocode.~\ref{alg:scale-mgmt}). 
Upon scaling out, the LBS updates the mapping in a reliable storage system and 
notifies each of the SGSs associated with this DAG to reinitialize the queuing delay windows 
so that we can observe the impact of our decision. The LBS makes the next
scaling decision only once the windows are filled up to avoid reacting to
transient changes in queuing delay.

\noindent\textbf{Scale In.} The LBS follows a similar process as described
above to decide if we need to dissociate an SGS from the DAG, with the only
difference being that we scale in if the scaling metric falls below the scale-in threshold (lines 9-10 in Pseudocode.~\ref{alg:scale-mgmt}). We remove the SGS that was added last from the pool of associated SGSs.
To avoid oscillations in the scaling process, we keep the scale-in threshold
well below the scale-out threshold.

\noindent\textbf{Scaling Metric.} Given the per-SGS queuing delay, in order to
calculate the scaling metric, we first compute a weighted sum of queuing
delays where we scale per-SGS queuing delay based on the number of proactively
allocated sandboxes that exist at the SGS (line 5 in Pseudocode.~\ref{alg:scale-mgmt}). Next, we normalize this weighted
sum by the available slack for the DAG (line 6 in Pseudocode.~\ref{alg:scale-mgmt}). Weighing the queuing delays
proportional to the number of sandboxes ensures that we give more (less)
importance to the SGS that handles more (less) requests of this DAG as the
sandboxes indicate what quantity of requests are handled by an SGS.
Normalizing by the available slack makes the scaling deadline-aware as it
scales-out more aggressively for latency-sensitive jobs compared to background
jobs as the former has less slack and queuing delays can lead to more missed
deadlines in comparison to the latter.

\subsubsection{How to do transparent scaling?}
\label{subsubsec:transparent_scaling}

When the LBS dynamically scales the SGSs associated with a DAG, we also need to
ensure that this does not have a negative impact on the requests. \name{}
achieves this by gradually scaling out and in rather than scaling instantly.

When scaling out, we associate an additional SGS with the DAG. However,
instantly sending requests to the new SGS will lead to these requests
experiencing sandbox allocation overheads. The LBS circumvents this issue by
gradually ramping up the newly added SGS in the following manner - (1) uses
\emph{lottery scheduling} to perform sandbox-aware routing among the various SGSs
where the number of tickets for each SGS correspond to the number of
proactive sandboxes it has setup for this DAG and (2) notifies the new SGS
to proactively allocate the average number of sandboxes present across the
active SGSs (calculated including the new SGS). We initialize the tickets for
the new SGS with a small value (say 1) so that requests go to it and this gets
updated as and when sandboxes are setup. Recall that the LBS knows about the
number of sandboxes allocated as they are piggy backed on the responses. The
system reaches steady-state once the required number of sandboxes have
been allocated.

Similarly, we also need to scale in gradually. An instant scale
in can result in overwhelming the reduced subset of SGSs. We solve this issue
by maintaining two lists of SGSs for a DAG - an {\em active list}
and a {\em removed list}. While scaling in, we remove the SGS from
the active list and place it in the removed list. During lottery scheduling, we still consider
SGSs in the removed list but scale down the lottery tickets given to such SGSs
by a {\em discount factor}. This ensures that the
subset is not overwhelmed and gradually removes the SGS.
\section{Implementation}
\label{sec:impl}

We built our prototype in Go (\textasciitilde$15K$ LOC). 
All the services  are implemented as multi-threaded processes. Our LBS 
has an HTTP front end to receive events that trigger the execution of the corresponding DAGs.
The SGS consists of the three loosely coupled modules - scheduler, estimator
and sandbox manager. All workers in the cluster have  
execution manager running as a daemon process. This daemon receives scheduling
requests from an SGS and places them in the corresponding core queues, and
also handles sandbox allocation/eviction requests. Currently, the prototype
supports docker containers as well as goroutines as sandbox environments. 
The external state store is responsible for keeping the SGS and LB
state and uses separate goroutines for handling requests. All the
communication between the different components happen using protocol buffers~\cite{protobuf}. We
integrate our prototype with Prometheus~\cite{prometheus} and
Grafana~\cite{grafana} for timely monitoring.
Next, we briefly describe the fault tolerance properties of our implementation.

\subsection{Fault Tolerance}
\label{sec:ft}

We assume the standard fail-stop model in which the \name{}'s services can crash at any point and that there
exists a failure detector that can immediately detect the failure.  

\noindent\textbf{Worker Failures.} When a worker fails, the corresponding SGS updates its
cluster view. Additionally, our per-DAG scaling strategy naturally adapts to
worker failures and limits the negative impacts on the incoming workload under
such situations. Specifically, when workers fail, the cumulative load that an
SGS can handle is reduced, and to meet deadlines, we
would ideally need to scale out. Since the scaling indicator is the queuing
delay, the LBS would observe an increased delay and scale out. 
Also, given that we evenly spread the proactive sandboxes, on
worker failure, incoming requests still benefit from proactive allocation on other workers.

\noindent\textbf{SGS and LB Failures.} \name{} maintains the state required by
the SGSs (e.g., proactive sandbox count, estimation state) and LB (per-DAG SGS
mapping) in a reliable external store. This ensures that a new instance can recover the
state from the store and continue execution. 
			
\section{Evaluation}
\label{sec:eval}

We evaluate the end-to-end benefits of \name{} on a 74-machine cluster deployed on CloudLab~\cite{cloudlab}
and compare against a baseline that reflects current state-of-the-art serverless platforms~\cite{openwhisk}.
We also carry out several microbenchmarks to delve deeper into \name{}'s benefits.

\subsection{Experimental Setup}
Our {\bf testbed} has 38 machines with 20 cores and  36 with
28 cores. All machines have 256GB memory and 10Gbps NIC. We partition the cluster 
to have 8 SGSs, each of which has a worker pool consisting of 8 machines. Each SGS runs on a separate machine. The setup uses a 
single load balancer to constitute the LBS.
We choose the {\em ScaleOutThreshold} to be 0.3 (\secref{subsec:sens_analysis}) and model the sandbox setup overheads for different DAGs to be in the range of 125 ms~\cite{firecracker}
to 400 ms, a conservative estimate given our measurements of overheads in downloading
code packages from S3 (\secref{subsec:real_world_study}).

\noindent{\bf Baseline Stack.} 
Our baseline uses a centralized scheduler (similar to~\cite{openwhisk}) where requests are processed in 
FIFO order. Also we {\em reactively} allocate sandboxes
and keep them in memory with a fixed inactivity timeout of 15 mins~\cite{openwhisk,lambda-cold,azure-cold}. 

\noindent{\bf Workload.} We consider four different classes of DAGs:
{\bf (i) C1} consists of DAGs that have a single function, short execution times and tight deadlines. These DAGs represent user-facing functions.
{\bf (ii) C2} consists of DAGs that have a single function, short execution times, and less strict deadlines. These DAGs represent non-critical user-facing functions (such as updating a metrics dashboard).
{\bf (iii) C3} consists of DAGs that have chained functions, medium execution times and
relatively strict deadlines compared to their execution times.  These DAGs represent more expensive
user-facing functions.
{\bf (iv) C4} consists of DAGs that have branched structures, high execution times and loose
deadlines. These DAGs represent background jobs that typically perform batch
execution~\cite{pywren}. We randomly sample execution time and slack details from the ranges
mentioned in Table~\ref{tab:sine_workload_details}.

\setlength\tabcolsep{4pt}
\begin{table}[t]
	\centering
	\begin{tabular}{ c | c | c | c | c | c}
		\hline
		&  \footnotesize\textbf{Avg. RPS} & \footnotesize\textbf{Amplitude} & \footnotesize\textbf{Period} & \footnotesize\textbf{Exec. Time} & \footnotesize\textbf{Slack} \\
		\hline
		\footnotesize C1 & \footnotesize [600,1200] & \footnotesize [100,800] & \footnotesize [10,20]s & \footnotesize [50-100]ms & \footnotesize [100,150] ms\\ 
		\hline
		\footnotesize C2 & \footnotesize [400,800] & \footnotesize [200,400] & \footnotesize [30,40]s & \footnotesize [100-200]ms & \footnotesize [300,500] ms\\
		\hline
		\footnotesize C3 & \footnotesize [500,1000] & \footnotesize [200,600] & \footnotesize [10,20]s & \footnotesize [250-400]ms & \footnotesize [200,300] ms\\
		\hline
		\footnotesize C4 & \footnotesize 200 & \footnotesize 0 & \footnotesize $\infty$ & \footnotesize [300-600]ms & \footnotesize [500,1000]ms\\
		\hline
	\end{tabular}
	\caption{Execution time and slack for various DAG classes (both Workloads). DAGs follow sinusoidal patterns in Workload 2 and we randomly sample the sinusoid pattern parameters from the range stated, depending on the class.}
	\label{tab:sine_workload_details}
\end{table}

\begin{figure*}[t!]
	\captionsetup[subfloat]{captionskip=-4pt}
	\centering 
	\subfloat[][]{%
		\includegraphics[width=0.25\textwidth]{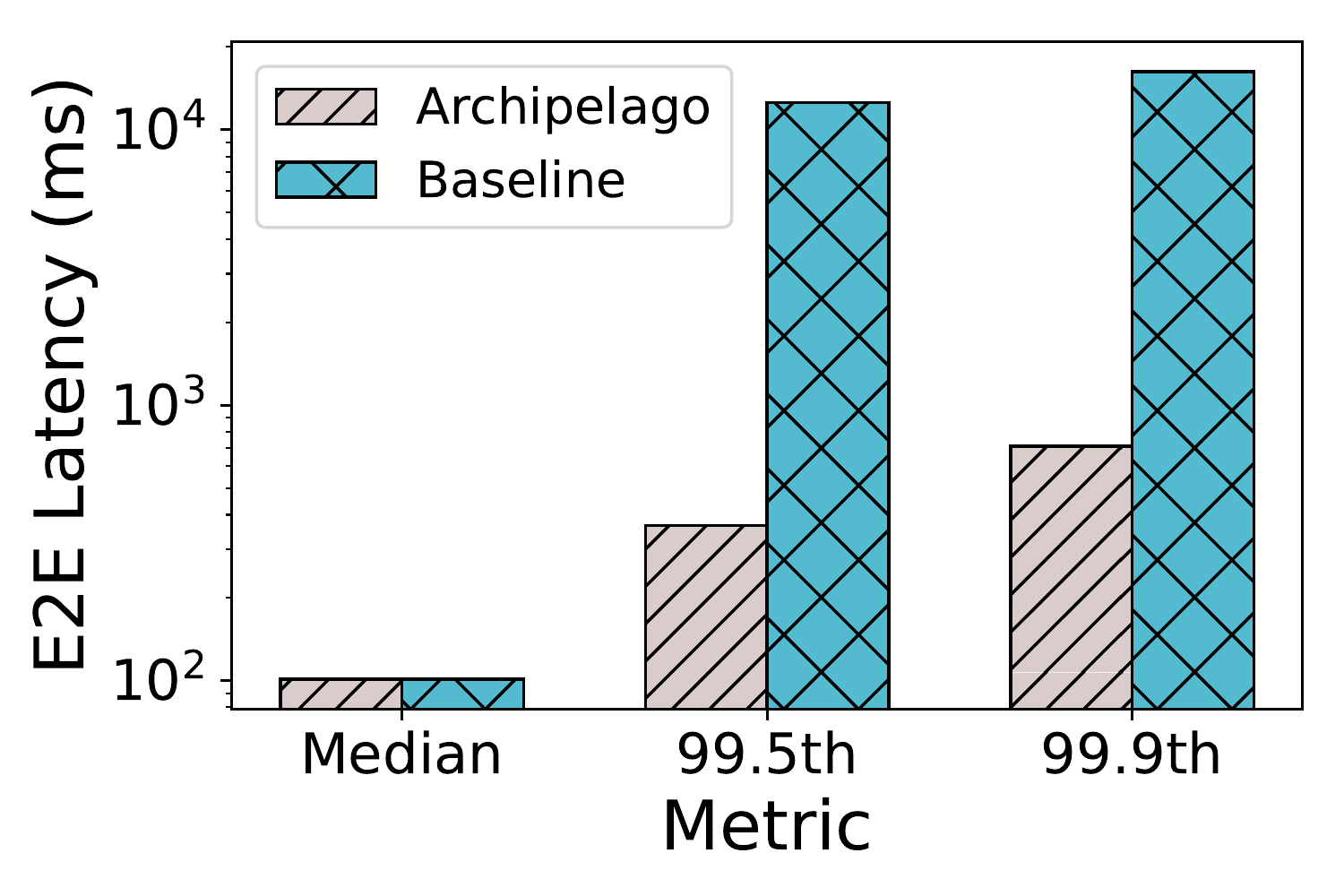}%
		\label{fig:poisson_e2e_latency}
	}
	\subfloat[][]{%
		\includegraphics[width=0.25\textwidth]{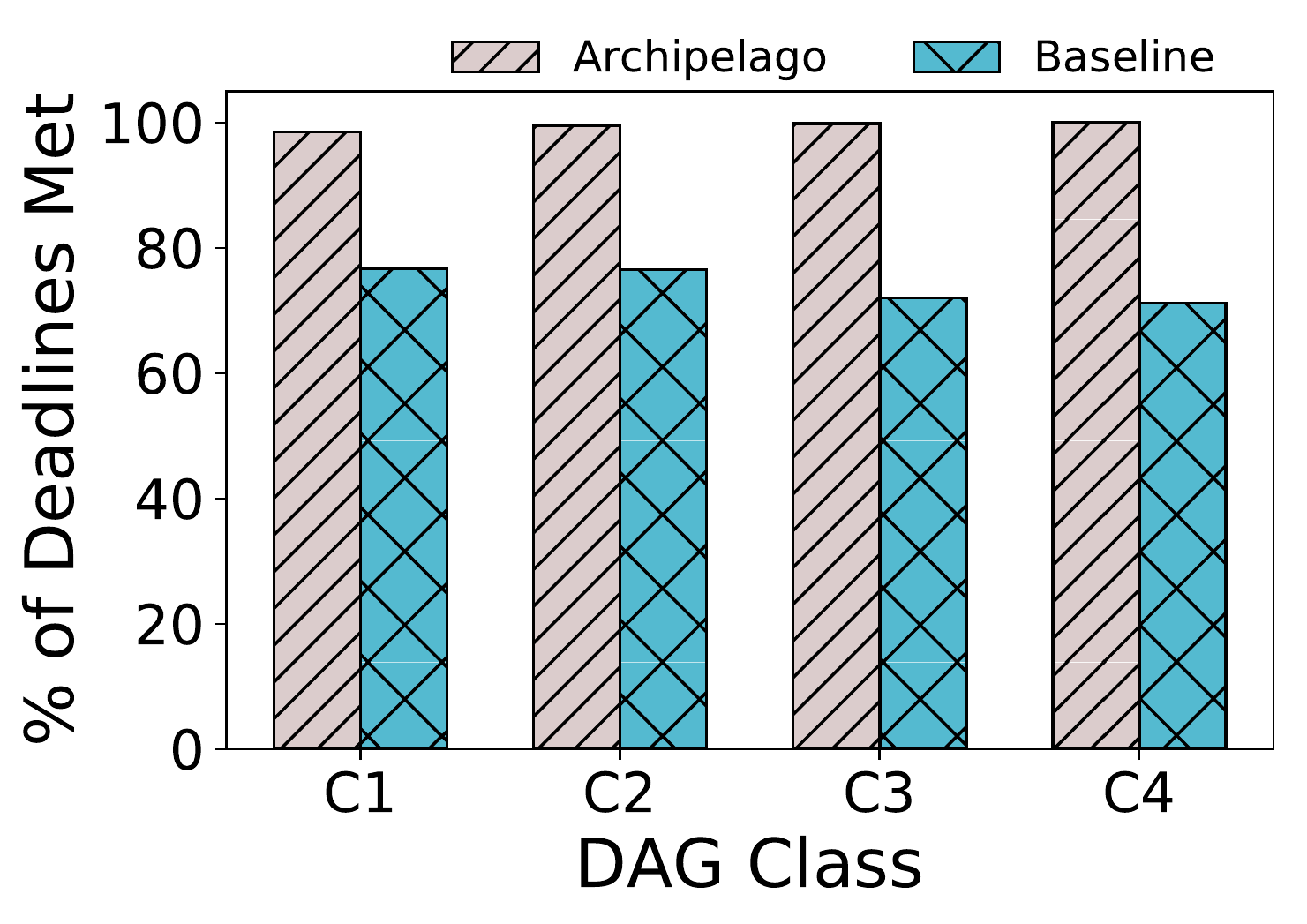}%
		\label{fig:poisson_deadlines}
	}
	\subfloat[][]{%
		\includegraphics[width=0.25\textwidth]{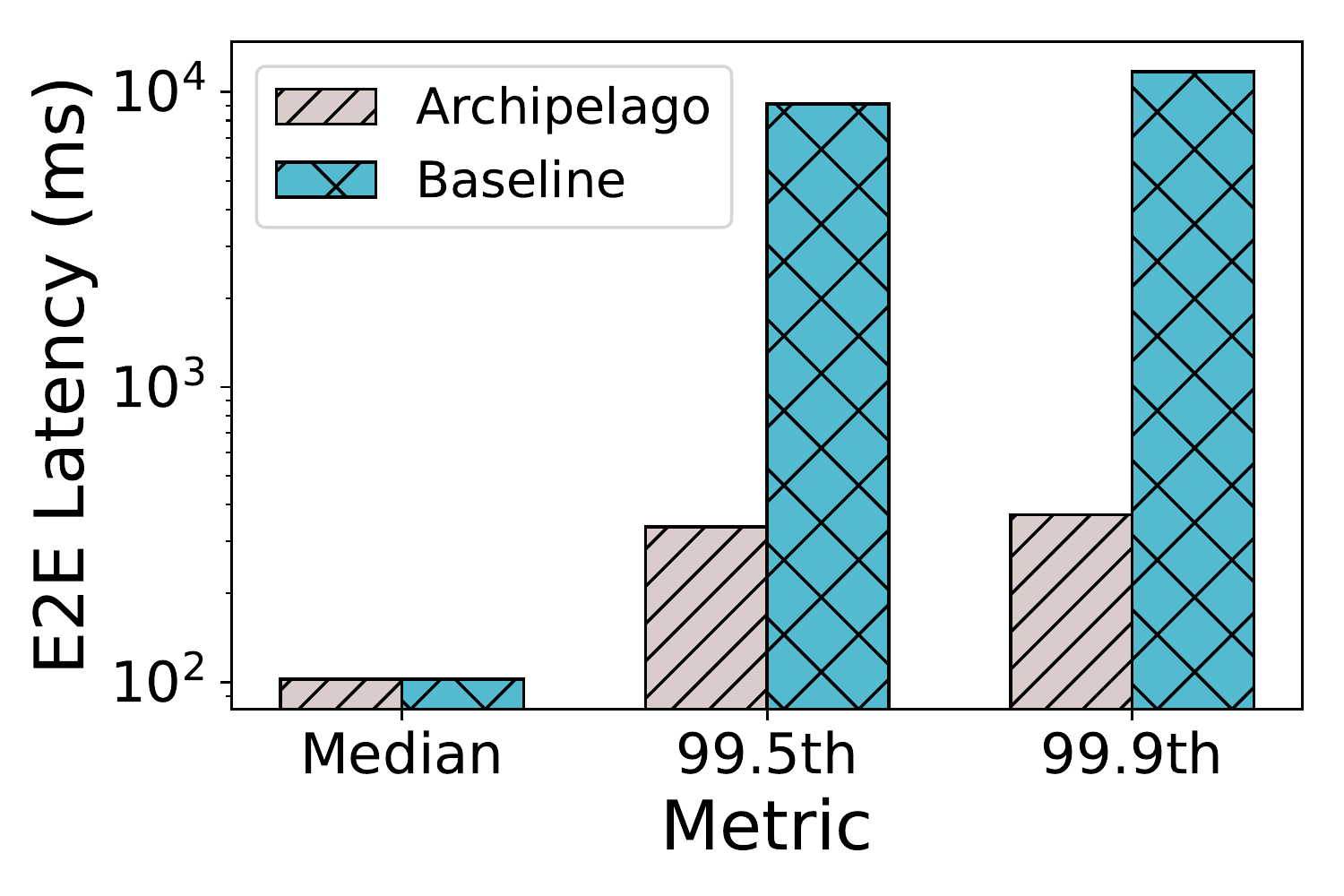}%
		\label{fig:macro_e2e_latency}
	}
	\subfloat[][]{%
		\includegraphics[width=0.25\textwidth]{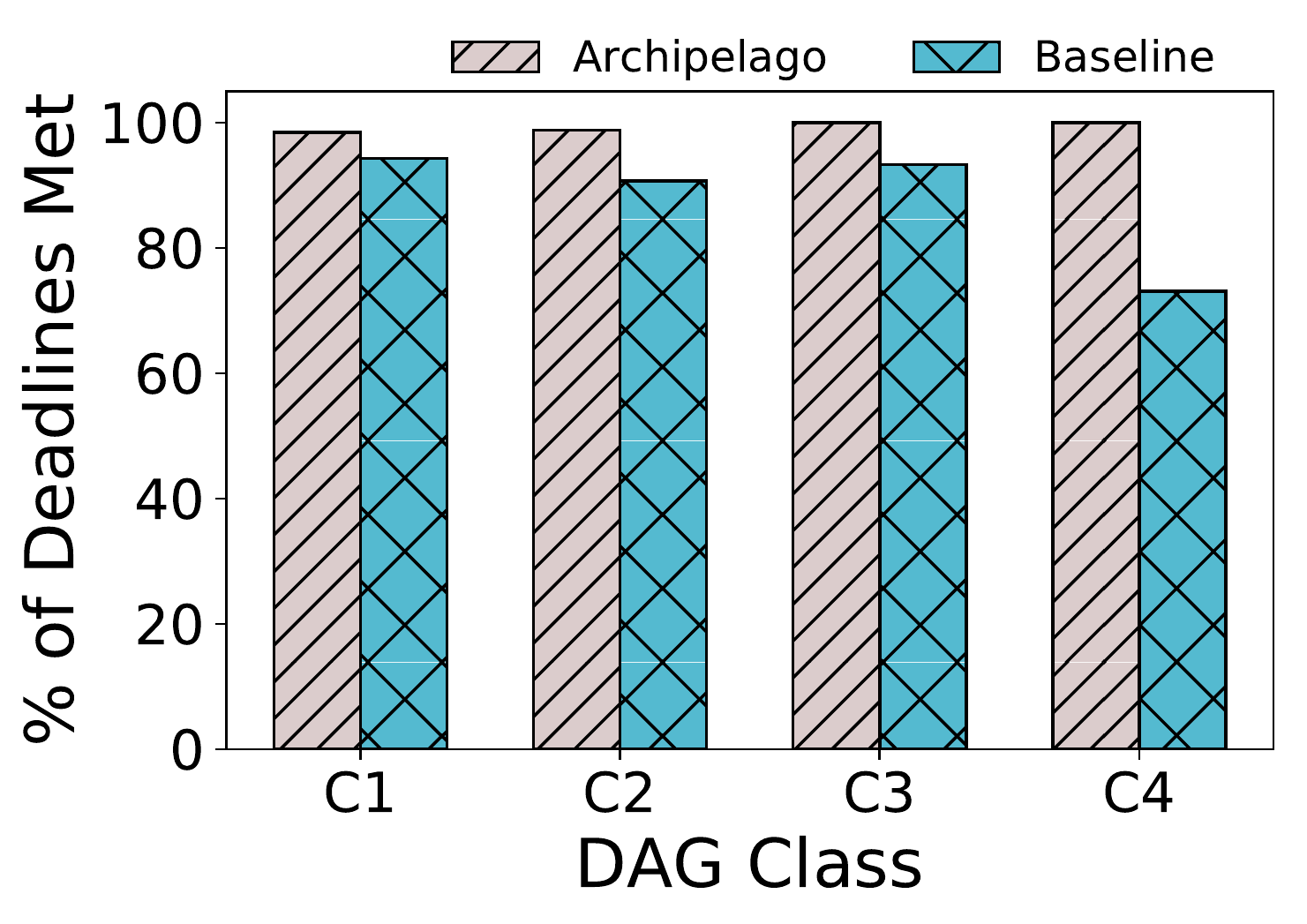}%
		\label{fig:macro_deadlines}
	} 	
	\vspace*{-2mm}
	\caption{ \footnotesize \name{} vs. baselines. (a) E2E Latency - Workload 1 (b) \% Deadlines Met - Workload 1 (c) E2E Latency - Workload 2 (d) \% Deadlines Met - Workload 2}
	\vspace*{-3mm}
\end{figure*}

We construct 2 workloads to model the arrival rate of requests
belonging to different classes.
For {\bf Workload 1}, we model the request arrival pattern to follow a Poisson distribution. 
For the classes C1-C4, we periodically (every second) sample the mean arrival rate
from an interval of 800-1200, 600-900, 600-800, 50-150 RPS respectively. For
{\bf Workload 2}, we model the request arrival pattern to follow a sinusoidal
distribution. The details are captured in Table
\ref{tab:sine_workload_details}. Both workloads keep the cluster CPU load between
\textasciitilde70\% to \textasciitilde110\%.

\noindent{\bf Metrics.} We use a variety of metrics to evaluate different components of the platform -
{\bf (i) End-to-end (E2E) latency} - represents the turn around time of a request. 
{\bf (ii) \% Deadlines Met} - the \% of requests that complete within their deadline.
{\bf (iii) Queuing Delay} - the time spent by a request in the queue before it is scheduled.
{\bf (iv) Cold Starts} - the number of requests that experience the overhead of sandbox allocation.

\subsection{Macrobenchmarks}

Figure~\ref{fig:poisson_e2e_latency} shows the end-to-end latencies for \name{} and the 
baseline for Workload 1. \name{} reduces the tail latencies (99.9\%-ile) by 20.83$\times$ over the 
baseline.  Additionally, in the steady state 
\name{} matches the performance of the baseline (50\%-ile).
Figure~\ref{fig:poisson_deadlines} shows that these tail latency violations lead to around 33\%
deadlines being missed by the baseline while \name{} misses only 0.76\% deadlines.   

We find that the high tail latencies incurred by the baseline come from requests getting
queued up while sandboxes are being reactively allocated. 
\name{} minimizes the number of cold starts by proactively 
allocating sandboxes and being deadline-aware (\secref{sources_improvement}). 
Similar results are observed for Workload 2 - \name{} reduces tail latencies by
35.97$\times$ over the 
baseline (Figure~\ref{fig:macro_e2e_latency}) and misses 0.98\% deadlines in comparison to 9.66\%
missed by the baseline (Figure~\ref{fig:macro_deadlines}). 

Additionally, in the context of baselines, we see that typically the classes
of DAGs that have a slower arrival rate miss more deadlines
(C4 misses more than others, C2 misses more than C1). Further analysis
indicates that DAGs with lower request rate tend to be
stuck behind requests from DAGs with higher request rate in the scheduling queue.
\name{} naturally mitigates this by using a queuing-aware scaling
indicator that triggers scale out to another SGS.

\subsubsection{Sources of Improvement}
\label{sources_improvement}

 \begin{figure}[t]
	\vspace*{-6mm}
	\captionsetup[subfloat]{captionskip=-2pt}
	\centering 
	\subfloat[]{%
 		\includegraphics[width=0.24\textwidth]{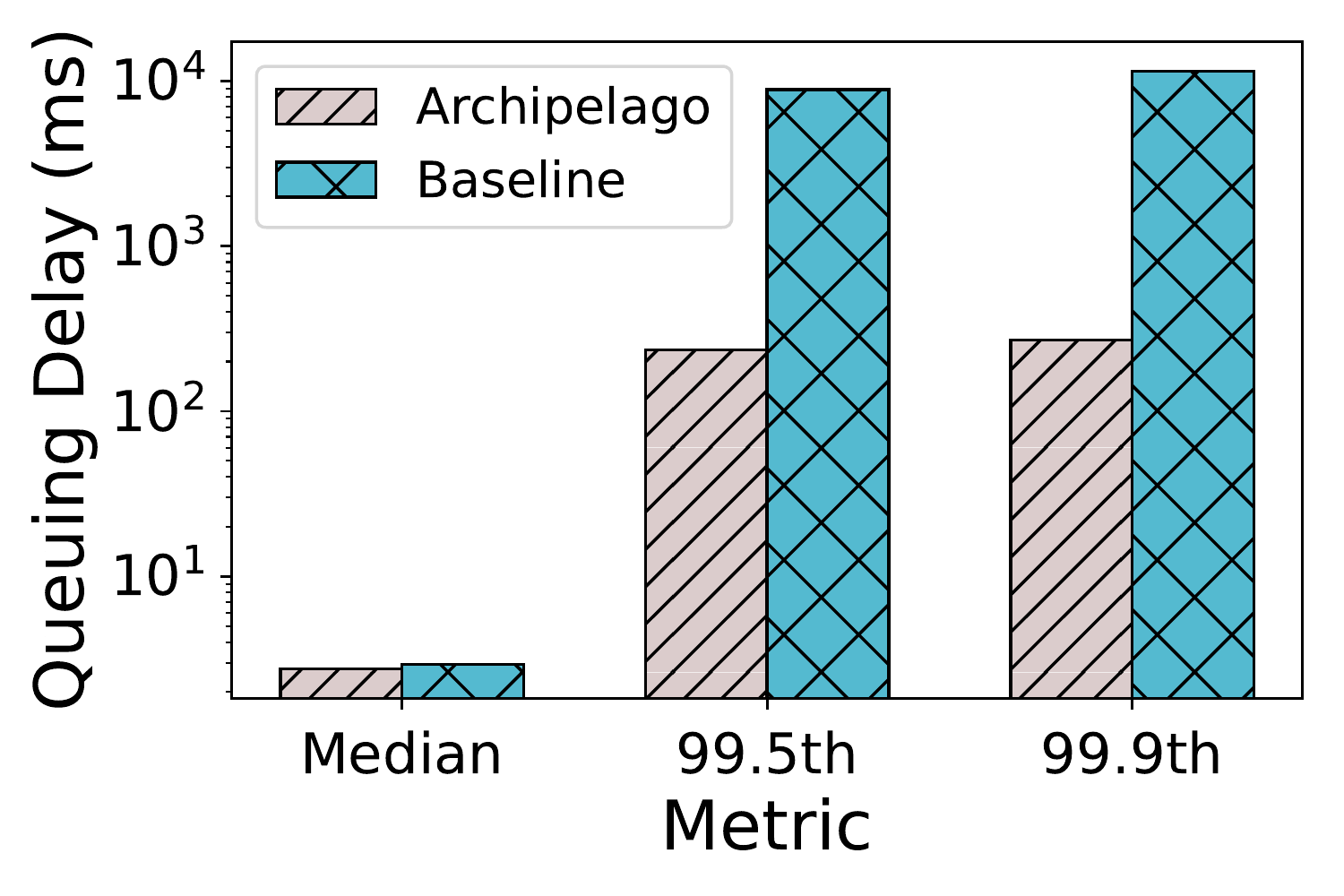}%
 		\label{fig:sources_improvement_queuing}
	}
	\subfloat[]{%
 		\includegraphics[width=0.25\textwidth]{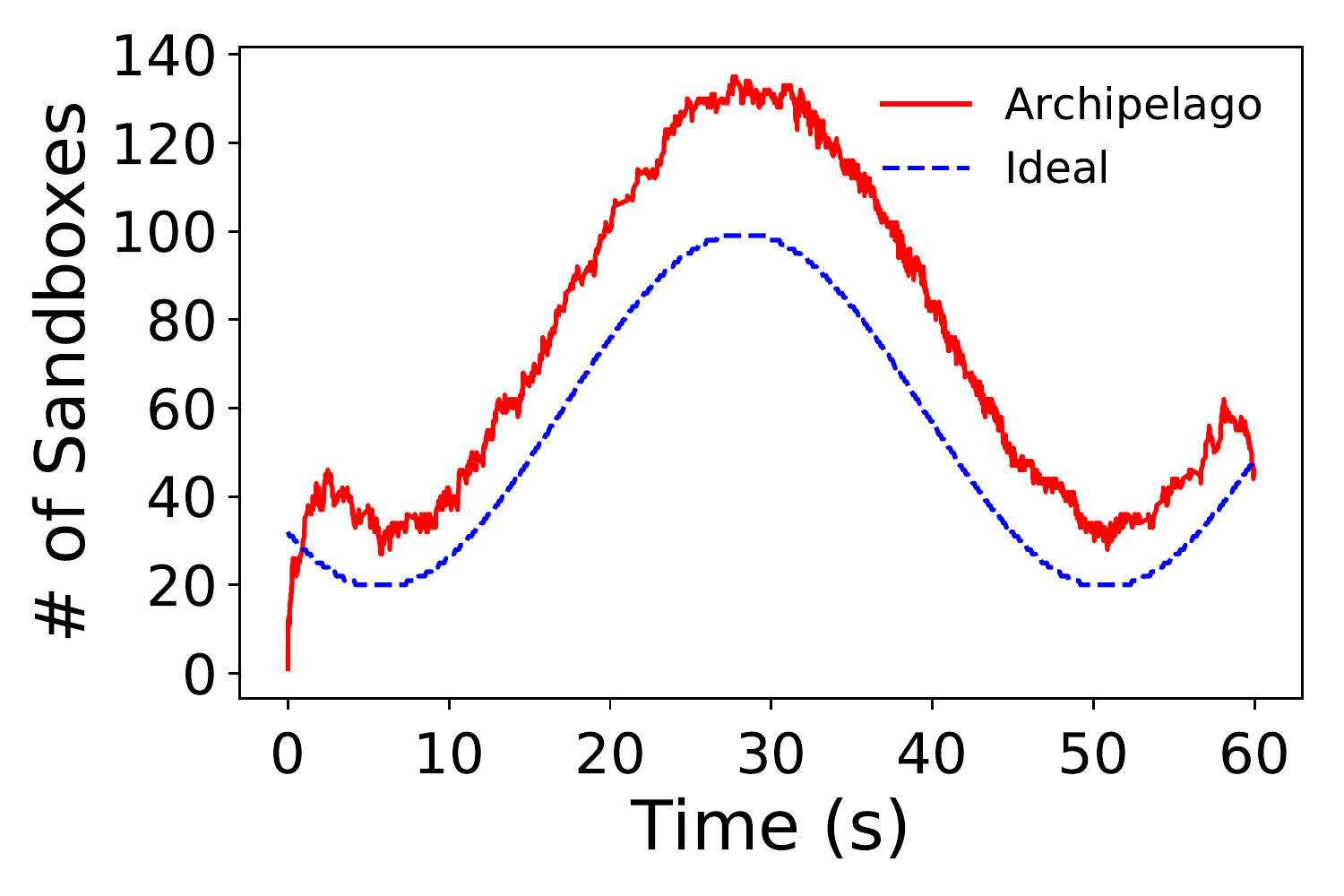}%
 		\label{fig:proactive_provisioning}
	}
    \vspace*{-2mm}
	\caption{ \footnotesize \name{} Workload 2 Sources of Improvement. (a) Queuing Delay (b) Proactive Allocation Vs Ideal Allocation}
	\vspace{-2mm}
\end{figure} 

We next analyze the sources of improvement
for the trends observed for Workload 2. We choose this workload to highlight
how \name{} behaves when the workload does not follow the Poisson arrival process assumed by our estimation logic.

\noindent{\bf Lower Queuing Delays.}
Figure~\ref{fig:sources_improvement_queuing}
shows that \name{} has lower queuing delays at an SGS. The tail queuing delay
for \name{} is 47.5$\times$ lower than the baseline. This is mainly due to - (i) LBS
performing sandbox-aware routing and (ii) SGS proactively allocating sandboxes
which ensures that requests do not spend additional time in the SGS queue
waiting for the allocation to finish. 

\noindent{\bf Fewer Cold Starts.} We see that \name{} overall incurs 24.38$\times$ 
fewer cold starts since sandboxes are proactively allocated in a workload-aware 
manner. In contrast, the baseline reactively allocates sandboxes leading to more cold starts.

\noindent{\bf Workload-aware proactive allocation.}
Figure~\ref{fig:proactive_provisioning} shows the number of proactively
allocated sandboxes for the C2 DAG.  We see that the SGSs' estimation is
able to closely follow the ideal number of sandboxes required. In the worst
case, \name{} allocates 37.4\% more sandboxes.
This is primarily because the SGS provisions sandboxes for the worst case load
to ensure requests do not incur cold starts (\secref{sss:sandbox_demand}). Additionally, there
are instances when an SGS allocates proactive sandboxes anticipating future
requests, but then the DAG scales out to another SGS due to contention at the
prior one. However, this is
not a concern since \name{} uses an isolated memory pool for proactive sandbox
allocation along with a workload-aware eviction policy.

\begin{figure*}[t!]
	\begin{minipage}[t]{0.28\textwidth}
		\includegraphics[width=\textwidth]{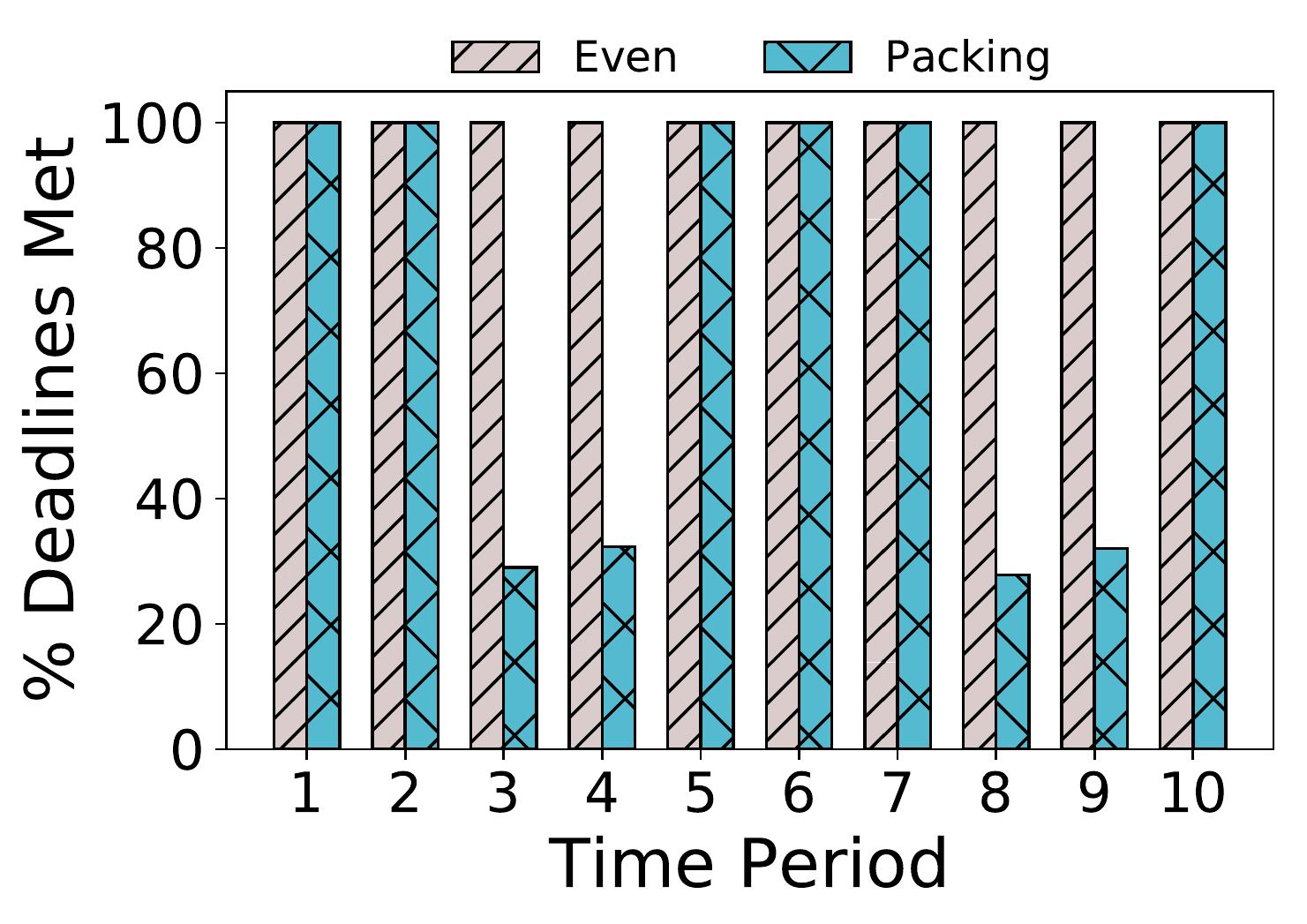} 
		\caption{Sandbox Placement - Even Vs. Packing}
		\label{fig:allocator_benefits}
	\end{minipage}
    \hspace{0.6cm}
	\begin{minipage}[t]{0.28\textwidth}
		\includegraphics[width=\textwidth]{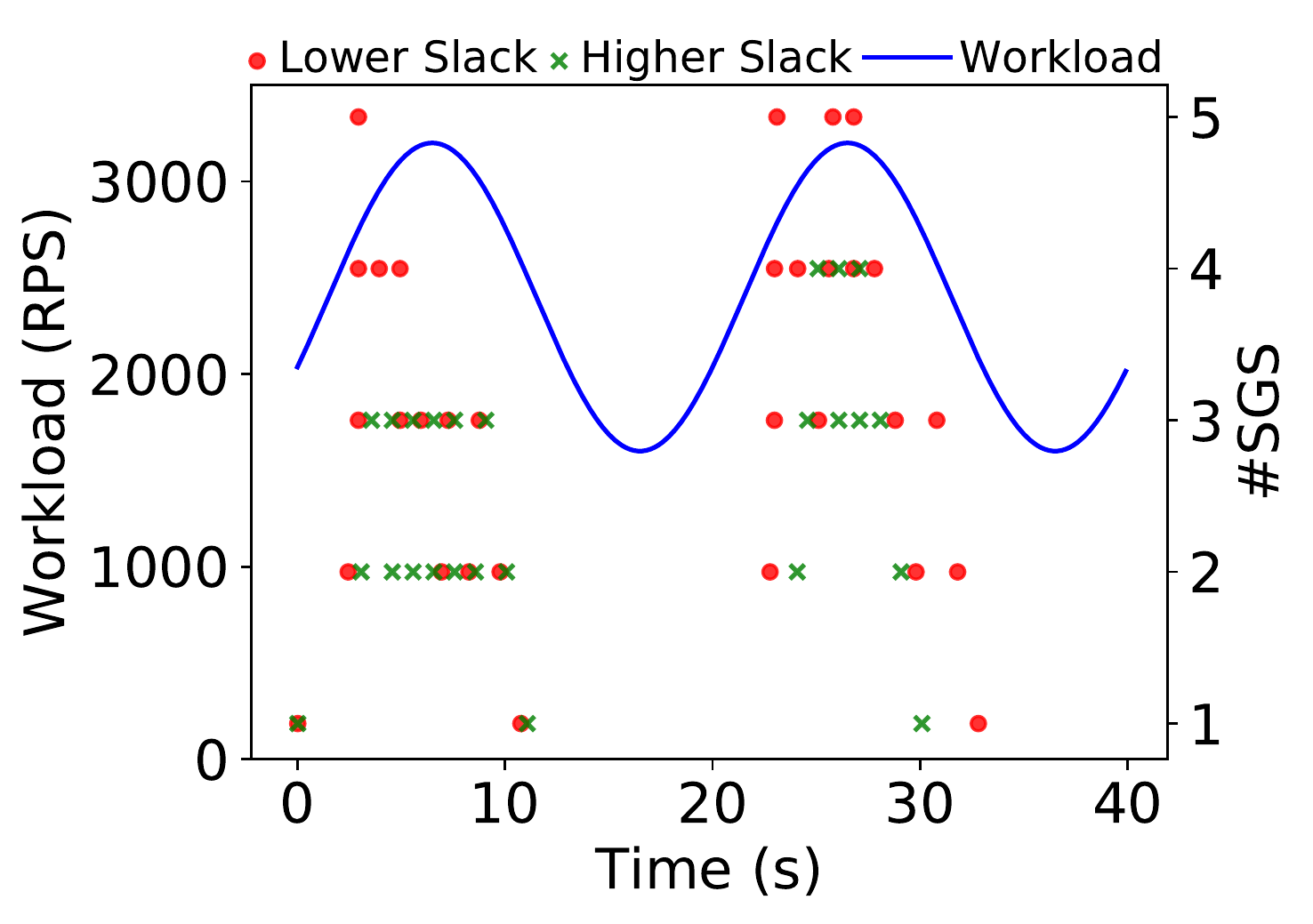}
		\caption{A DAG with lower slack scales-out more than a DAG with higher slack}
		\label{fig:slack_aware_scaling}
	\end{minipage}%
    \hspace{0.6cm}
	\begin{minipage}[t]{0.28\textwidth}
		\includegraphics[width=\textwidth]{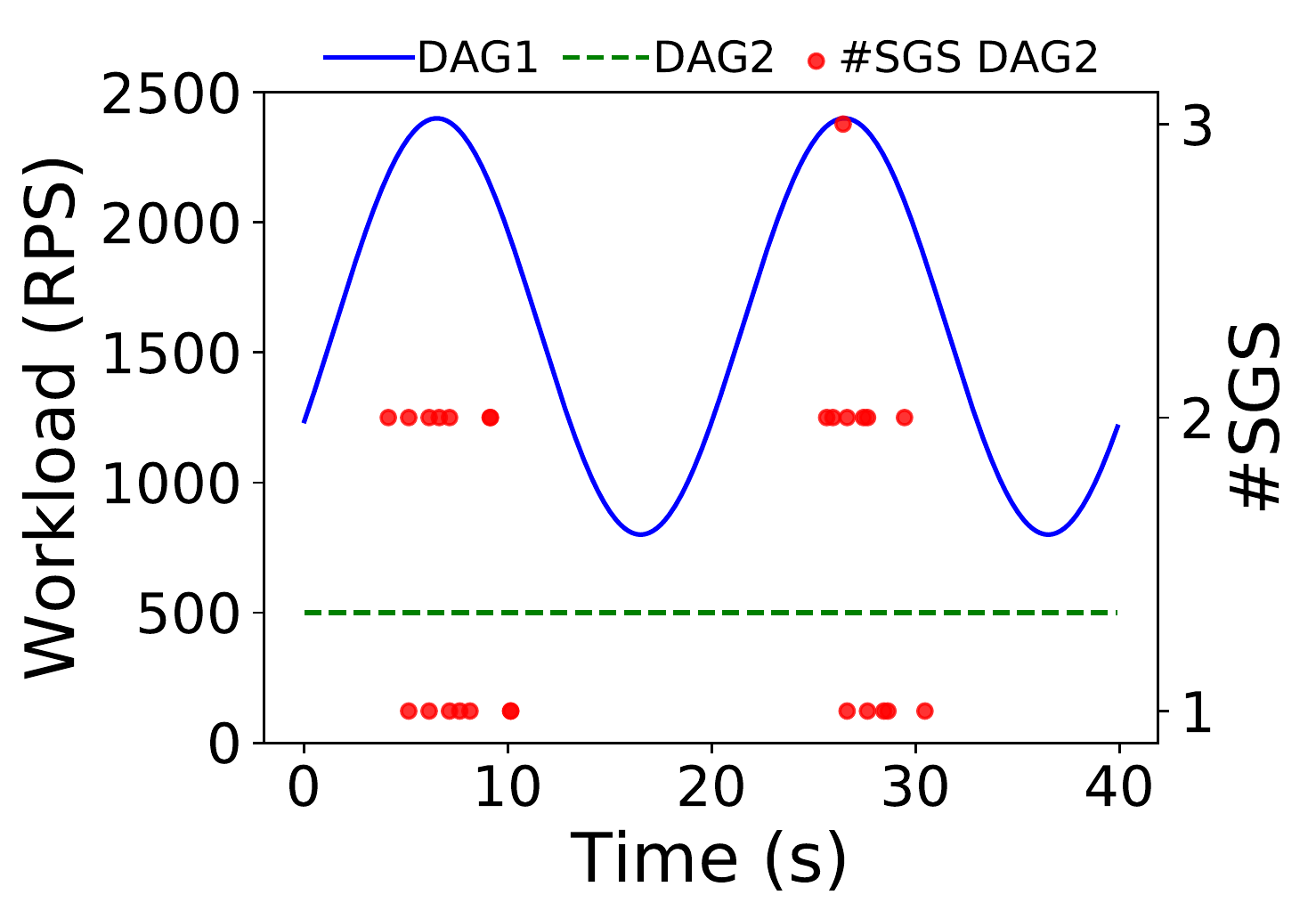}
		\caption{Contention from a bursty DAG (DAG1) causes DAG2 to scale-out}
		\label{fig:cross_dag_scaleout}
	\end{minipage}
\vspace{-0.3cm}
\end{figure*}

\subsection{Microbenchmarks}
To further delve into the benefits of \name{}, we run several
microbenchmarks at a smaller scale, with 1 LB, one or more SGSs, and
each SGS having 10 workers.  We use synthetic workloads that stress
specific components of the stack.

\subsubsection{SGS Sandbox Management}
\label{subsubsec:eval_sandbox_management}

We study the effectiveness of the sandbox placement and eviction 
against alternative strategies using one SGS.

\noindent{\bf Evenly spreading sandboxes.} We compare our approach of evenly spreading
sandboxes across the worker pool to an alternative where the SGS packs sandboxes on the same worker.
We choose a 
workload with a single DAG where the request arrival follows a sinusoidal distribution
with an average RPS of 1200, amplitude of 600, and a 20s period. 

Given that both approaches see the same workload, the number of proactive sandboxes
allocated are the same. However, we observe
(see Figure~\ref{fig:allocator_benefits}) that the packing approach leads to
\textasciitilde70\% deadlines not being met during intervals of increased load
(intervals 3-4, 8-9). This does not happen when sandboxes are evenly spread.
This is primarily because in case of packing, the sandboxes are available on
a smaller fraction of workers, and at increased load, requests gets scheduled
on workers that do not have proactively allocated sandboxes available, leading to missed deadlines.
In contrast, even placement of sandboxes offers better statistical multiplexing
resulting in better handling of bursts.

\noindent{\bf Benefits of workload-aware hard eviction.} We compare our approach of fair eviction with LRU (\secref{sss:eviction}).
We choose a workload that
consists of 2 DAGs - one that has constant request rate of 200 RPS and another
one that has an on/off pattern with 100 RPS. We have configured the
proactively memory pool to be low so that it causes hard eviction. We observe
that LRU has a higher tail latency by $4.62X$ in comparison to fair eviction. This is primarily due to LRU optimizing for the short-term without taking into
account the sandbox demand, which \name{} does. Specifically, we observe that
during the off-period, using LRU causes all sandboxes of the second DAG to be hard
evicted leading to additional sandbox setup overheads
during the next on period.

\subsubsection{LBS Scaling Strategy} 
\label{subsubsec:lbs_micro}

We now evaluate the various aspects of the scaling strategy adopted by the
LBS using 5 SGSs with 10 workers each.

\noindent{\bf Benefits of gradual scale-out.} \name{} gradually scales-out the
number of SGSs for a given DAG using lottery scheduling
(\secref{subsubsec:transparent_scaling}). We evaluate the benefit of this
against a policy where scale-out happens instantly, which leads to
LBS routing requests in a round-robin fashion among the
SGSs. 
We choose a 
workload with a single DAG wherein the request arrival follows a sinusoidal distribution
with an average RPS of 800, amplitude of 600, and a 100s period (elongated period to capture a snapshot of the scale-out benefits). 

We observe 1.5$\times$ higher tail latencies with instant scale-out.
This is because when a new SGS is added for a DAG, the LBS immediately starts
routing requests to it, without taking into account the number of available
sandboxes.

\noindent{\bf Deadline-aware per-DAG scale-out.}
\name{}'s per-DAG scaling metric accounts for the amount of slack in the DAG.
To study the effect of this, we consider 2 DAGs, both having an execution time of
100 ms. However, one DAG has a slack of 50 ms while the other has a slack of
200 ms. We assume a workload where requests arrive with the same sinusoidal
distribution (see Figure~\ref{fig:slack_aware_scaling} for workload).

From Fig~\ref{fig:slack_aware_scaling}, we observe that the DAG with smaller
slack scales-up to more SGSs than the DAG with higher slack 
(e.g., smaller slack DAG scales out to 4 while the larger slack DAG scales out to 3 in the 20-30s interval). 
This shows the
benefits of having a deadline-aware scaling metric which can help latency-sensitive foreground apps over
background apps.

\noindent{\bf Contention-aware per-DAG scale-out.}
Since DAGs from multiple users are multiplexed across the
same cluster, it is important to ensure that one DAG does not
suffer due to increased request rates of another DAG.
To evaluate this, we consider 2 DAGs - one that is bursty and follows a sinusoidal
distribution and another that has a low, constant request rate. The request rate of the second DAG
is set such that it requires only a single SGS if it is the only DAG utilizing the cluster (see
Figure~\ref{fig:cross_dag_scaleout} for workload).

When the second DAG experiences contention for the cluster due to the bursty nature of the first
DAG, we observe from Figure \ref{fig:cross_dag_scaleout} that the LBS is able to handle this
by scaling-out the second DAG to another SGS (e.g., at \textasciitilde
5s). We also notice that the LBS scales-down once the contention reduces (e.g., at
\textasciitilde 17s). This is possible since we co-design the LBS and SS layers,
allowing us to observe the contention at each SGS and appropriately scale in a deadline-aware
manner.

 \begin{figure}[t]
	\vspace*{-6mm}
	\captionsetup[subfloat]{captionskip=-3pt}
	\centering 
	\subfloat[]{%
 		\includegraphics[width=0.24\textwidth]{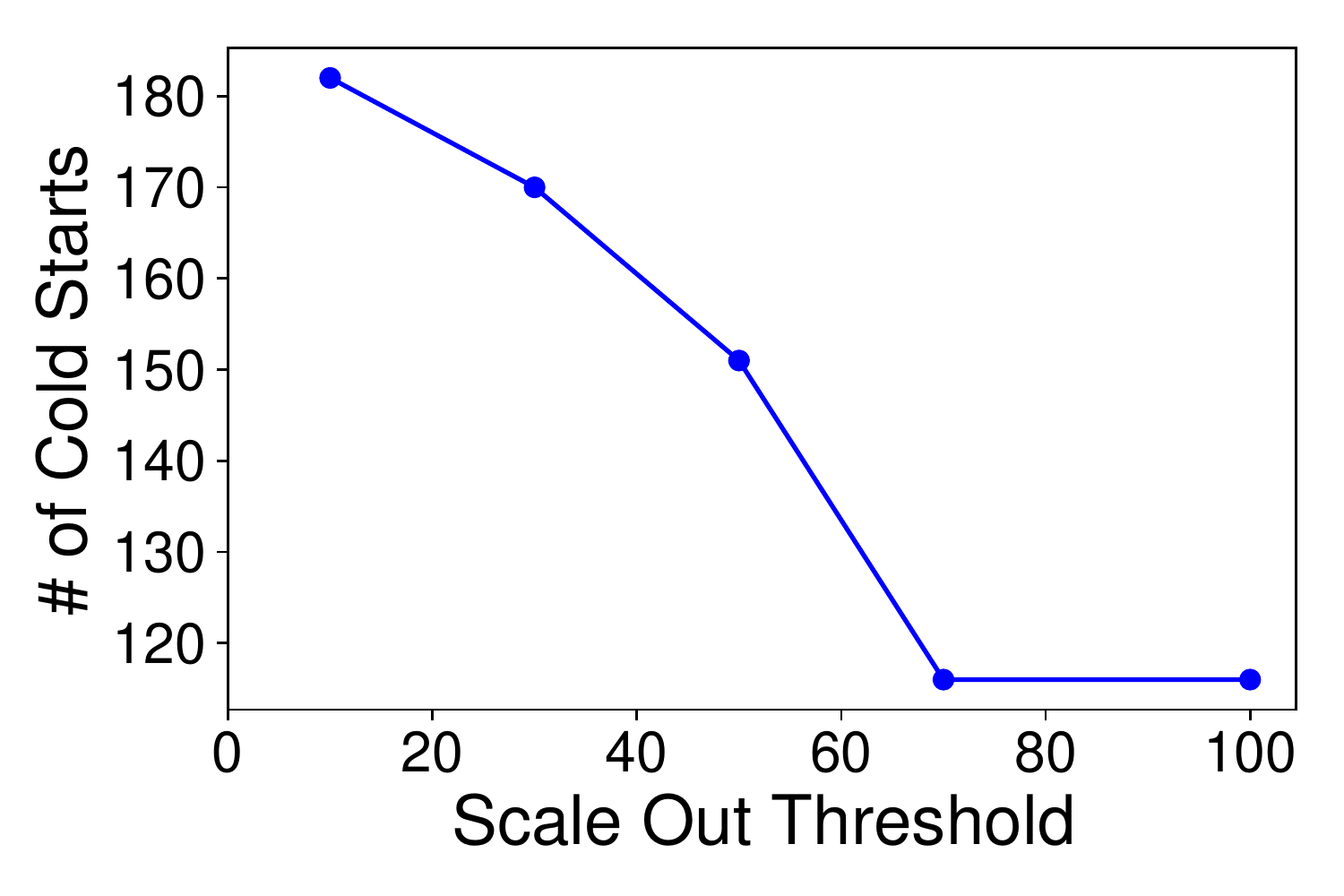}%
		\label{fig:lb_sens_cold_starts}
	}
	\subfloat[]{%
		\includegraphics[width=0.24\textwidth]{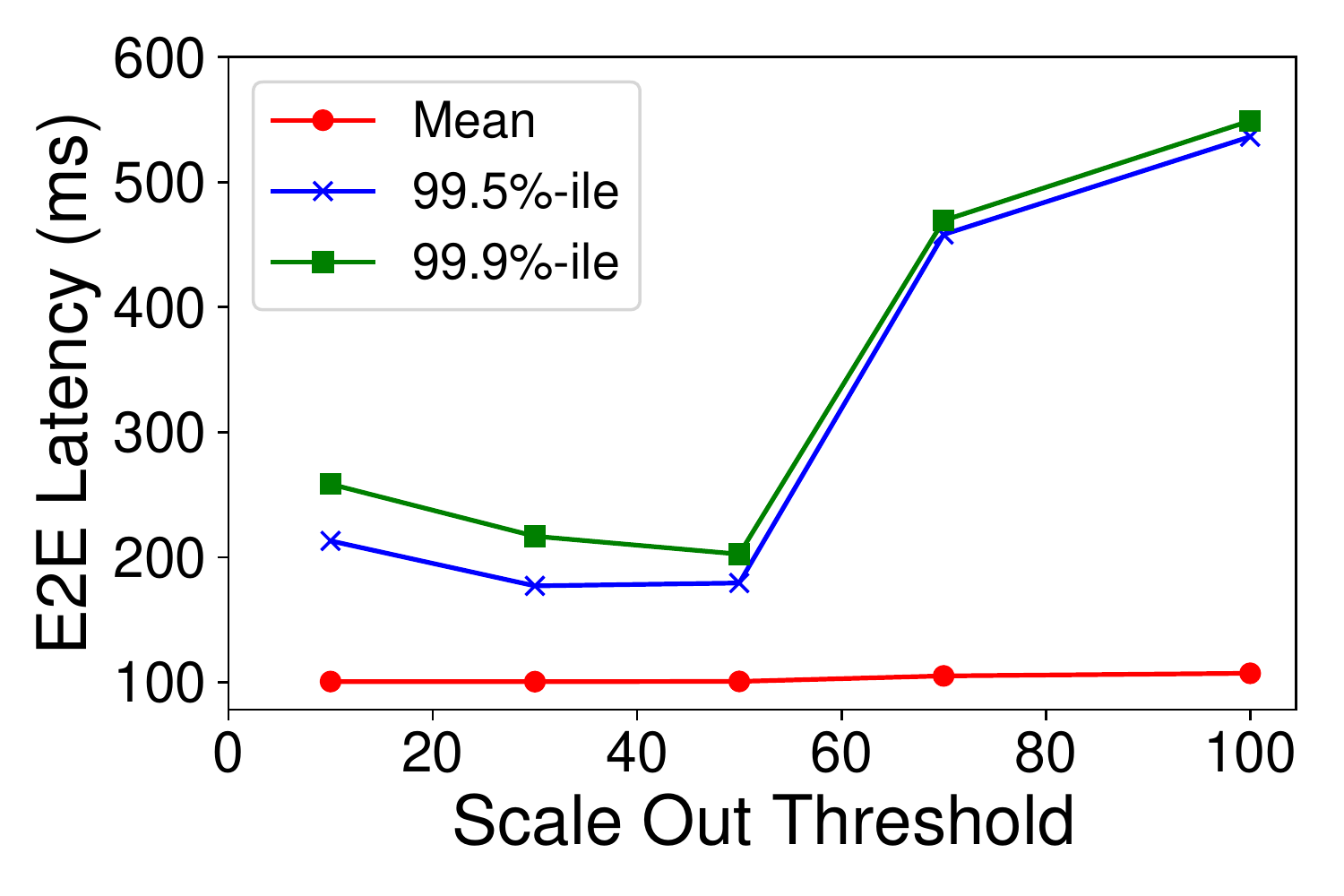}
		\label{fig:lb_sens_e2e_latency}
	}
    \vspace*{-3mm}
	\caption{ \footnotesize Scale Out Threshold Vs. (a) Cold Starts and (b) E2E Latency}
\end{figure} 

\subsection{System Overheads}

Since \name{} aims to provide low latency scheduling, we present
some of the overheads that arise in the critical path of request execution.
From our macrobenchmarks, we notice that the median (99\%-ile) per request
overhead added by the LBS to decide where to route is 190$\mu$s (212$\mu$s).
Scheduling decisions at SGS added an additional median (99\%-ile)
overhead of 241$\mu$s (342$\mu$s) per request. We also measure the time taken to scale-out at the LBS as well as time to make an
estimation decision. Neither of  these happen in the critical path, but help determine the
robustness of the system. Scale-out takes a median (99\%-ile) time of 128$\mu$s (197$\mu$s).
Estimations at an SGS take a median (99\%-ile) time of 879$\mu$s (1352$\mu$s).

\subsection{Sensitivity Analysis}
\label{subsec:sens_analysis}

{\bf Scale Out Threshold (SOT).} Lower values of SOT mean that the LBS scales-out more aggressively. This would
result in more frequent scale-outs amounting to a greater number of cold starts as seen in Figure
\ref{fig:lb_sens_cold_starts}. On the other hand, aggressive scale-out helps keep queuing delays low
in comparison to a passive scale-out strategy. Thus, we observe a trade-off between managing queuing
delays and the number of cold starts. From Figure \ref{fig:lb_sens_e2e_latency}, we observe that -
(i) At very low {\em SOT} values, the high number of cold starts negatively impacts
the tail latency (ii) At higher {\em SOT} values, higher queuing delays
negatively impacts the tail latency. A cluster operator can thus configure the {\em SOT} based on knowledge of the workload
and the sandbox setup overheads. Based on the above observed values, we choose a {\em
SOT} of 0.3 for our experiments.

\noindent{\bf SGS Size.} Given a fixed number of workers, what should be ideal size of the worker pool
under a single SGS? To study this, we consider a setup consisting of 20
workers. We consider 4 ways in which the cluster can be partitioned - (i) 20 SGSs, 1 worker each 
(ii) 10 SGSs, 2 workers each (iii) 5 SGSs, 4 workers each (iv) 1 SGS, 20 workers each.
We choose a workload with a single DAG wherein the request arrival follows a sinusoidal distribution
with an average RPS of 600, amplitude of 400, with a period of 20 seconds.

\begin{figure}[t!]
	\vspace*{-6mm}
	\captionsetup[subfloat]{captionskip=-3pt}
	\centering
	\subfloat[]{\includegraphics[width=0.49\columnwidth]{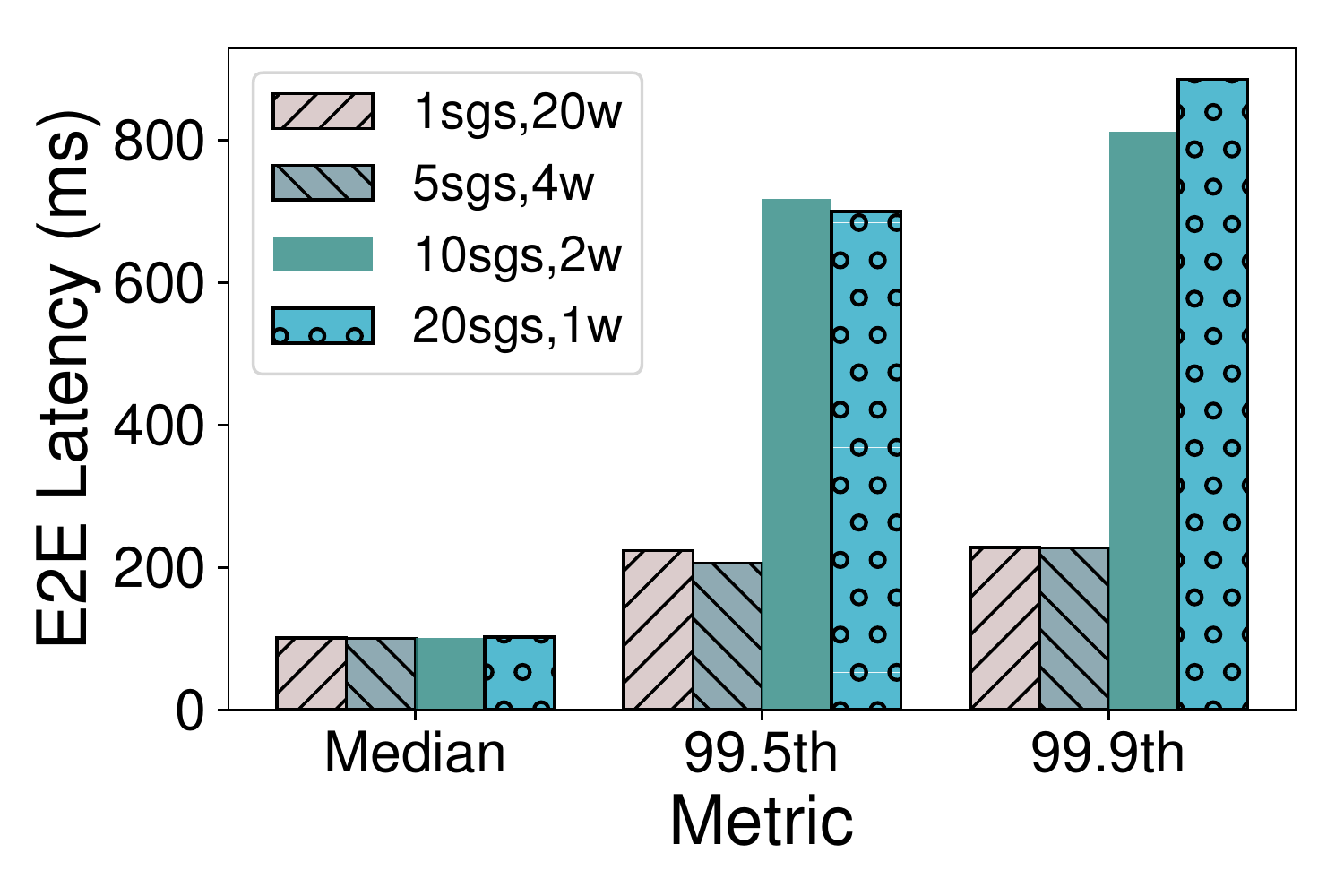}} \label{sgs_sizing_stats.pdf}
	\subfloat[]{\includegraphics[width=0.49\columnwidth]{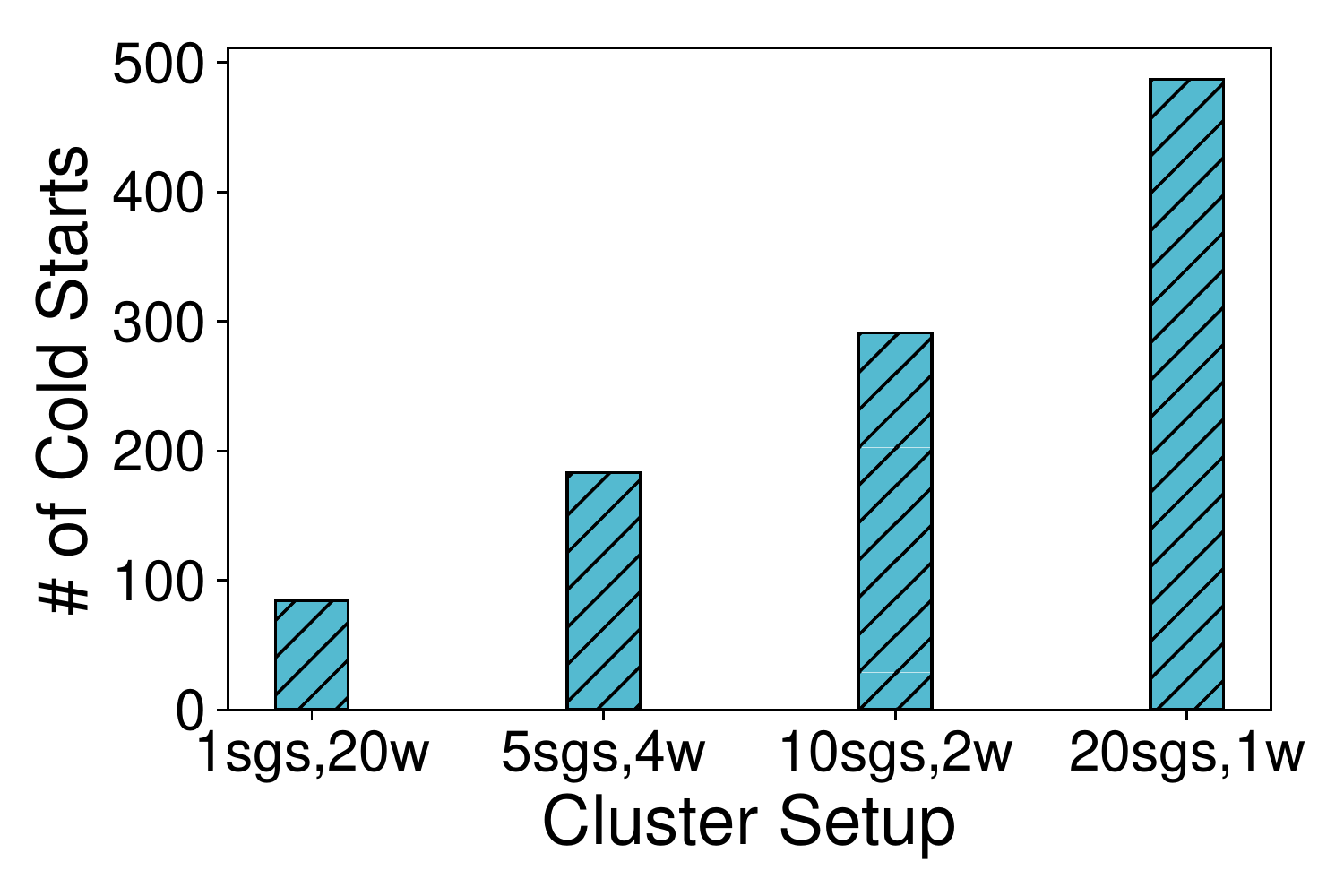}} \label{fig:sgs_size_cold_starts}
	\caption{\label{fig:sgs_sizing}
		Comparison of (a) E2E latencies and (b) Cold starts for different cluster configurations
	}
    \vspace{-3mm}
\end{figure}

We observe that fine-grained partitioning leads to \textasciitilde4$\times$
higher tail latencies (Figure~\ref{fig:sgs_sizing}(a)). This is because the LB
would need to scale-out more often for each DAG leading to an increased number
of cold starts in comparison to when there is no need to scale out as seen in
Figure~\ref{fig:sgs_sizing}(b).

However, having too many workers under an SGS can lead to scheduling overhead
becoming a significant contributor to the queuing delay. 
If this happens, then the LBS would unnecessarily scale out leading to workers under the initial SGS
being under-utilized. 
For functions with 50ms slack, we observed in our
testbed, that beyond 64 machines, we were unnecessarily scaling out leading to
workers being underutilized.
			
\section{Related Work}
\label{sec:related}

\noindent\textbf{Serverless Characterization.}~\cite{serverlessNF} looks at how network
intensive applications run on serverless platforms whereas ~\cite{pocket-workshop, pocket, locus}
characterize the storage requirements of serverless applications.~\cite{serverless-peek} conducted a
large measurement study to understand performance, resource management as well isolation in
serverless platforms. Similarly, ~\cite{4-benchmark} also conducted measurements on the public
offerings of serverless frameworks. To the best of our knowledge, no prior works have characterized  real world
serverless applications.    

\noindent\textbf{Sandbox Overhead Reduction.} ~\cite{sock} reduces the start up times of functions
in OpenLambda~\cite{openlambda} through caching Python runtimes and packages, and uses low-latency isolation
primitives.~\cite{micro} advocates for the usage of language-based isolation instead of using
traditional virtualization techniques.~\cite{sand} proposes a two level isolation wherein functions
of the same application run within the same container as separate processes.~\cite{agile} identifies
that the container networking setup takes significant time and pre-creates such resources to
overcome the overhead, and dynamically binds to a container. All these works are complementary with
\name{}'s efforts of reducing the impact of sandbox setup overheads.

\noindent\textbf{Scheduling Architectures.} We now discuss scheduling architectures other that those compared to earlier (i.e.,~\cite{sparrow,ray}). Borg~\cite{borg} uses random sampling while calculating scores and thus trades off scheduling optimality for scalability.
Omega~\cite{omega} uses multiple parallel schedulers but trades off scheduling predictability for
scalability due to the overheads involved in resolving conflicts which would happen often in our
setting due to the resources being held for short durations. While Apollo~\cite{apollo} tries to
reduce the frequency of conflicts by collecting cluster load periodically and feeding this to
individual job schedulers, it does not allow for diverse applications to share the cluster as it
makes the assumption that there are either latency sensitive tasks with guarantees or opportunistic
tasks with no guarantees. In \name{}, we can accommodate various kinds of tasks and meet
deadlines for all of them. Mercury~\cite{mercury} is a hybrid scheduler that makes high-quality assignment
for long tasks but the short tasks are scheduled in a distributed manner and can be preempted anytime
leading to sub-optimal placement for the shorter tasks.
						
\section{Conclusion}
\label{sec:conclusion}

In this paper, we consider the problem of ensuring low latency function execution in serverless settings, an important problem that has not received attention. Our system, \name{},  meets this goal using the following combination of simple but effective, scalable techniques - (a) partitioning the cluster into (semi-global scheduler, worker pool) pairs, (b) performing deadline-aware scheduling and proactive sandbox allocation, and (c) sandbox-aware routing with automatic scaling. Our evaluation shows that \name{} meets the deadlines for more than 99\% of realistic application request workloads, and reduces tail latencies by up to \textasciitilde$36X$ compared to state-of-the-art.

\label{EndOfPaper}
{
	\bibliography{archipelago}


\begin{thebibliography}{49}


\ifx \showCODEN    \undefined \def \showCODEN     #1{\unskip}     \fi
\ifx \showDOI      \undefined \def \showDOI       #1{#1}\fi
\ifx \showISBNx    \undefined \def \showISBNx     #1{\unskip}     \fi
\ifx \showISBNxiii \undefined \def \showISBNxiii  #1{\unskip}     \fi
\ifx \showISSN     \undefined \def \showISSN      #1{\unskip}     \fi
\ifx \showLCCN     \undefined \def \showLCCN      #1{\unskip}     \fi
\ifx \shownote     \undefined \def \shownote      #1{#1}          \fi
\ifx \showarticletitle \undefined \def \showarticletitle #1{#1}   \fi
\ifx \showURL      \undefined \def \showURL       {\relax}        \fi
\providecommand\bibfield[2]{#2}
\providecommand\bibinfo[2]{#2}
\providecommand\natexlab[1]{#1}
\providecommand\showeprint[2][]{arXiv:#2}

\bibitem[\protect\citeauthoryear{??}{Azu}{2017}]%
        {AzureFaas}
 \bibinfo{year}{2017}\natexlab{}.
\newblock \bibinfo{title}{{{Azure Functions}}}.
\newblock \bibinfo{howpublished}{\url{https://functions.azure.com}}.
\newblock


\bibitem[\protect\citeauthoryear{??}{GCF}{2017}]%
        {GCF}
 \bibinfo{year}{2017}\natexlab{}.
\newblock \bibinfo{title}{{{Google Cloud Functions}}}.
\newblock \bibinfo{howpublished}{\url{https://cloud.google.com/functions}}.
\newblock


\bibitem[\protect\citeauthoryear{??}{ope}{2017}]%
        {openwhisk}
 \bibinfo{year}{2017}\natexlab{}.
\newblock \bibinfo{title}{{{IBM Bluemix Openwhisk}}}.
\newblock
  \bibinfo{howpublished}{\url{https://www.ibm.com/cloud-computing/bluemix/openwhisk}}.
\newblock


\bibitem[\protect\citeauthoryear{??}{sla}{2019}]%
        {slack}
 \bibinfo{year}{2019}\natexlab{}.
\newblock \bibinfo{title}{{{Amazon Simple Notification Service}}}.
\newblock \bibinfo{howpublished}{\url{https://slack.com/}}.
\newblock


\bibitem[\protect\citeauthoryear{??}{s3}{2019}]%
        {s3}
 \bibinfo{year}{2019}\natexlab{}.
\newblock \bibinfo{title}{{{Amazon Simple Storage Service}}}.
\newblock \bibinfo{howpublished}{\url{http://aws.amazon.com/s3}}.
\newblock


\bibitem[\protect\citeauthoryear{??}{aws}{2019a}]%
        {aws}
 \bibinfo{year}{2019}\natexlab{a}.
\newblock \bibinfo{title}{{{Amazon Web Services}}}.
\newblock \bibinfo{howpublished}{\url{https://aws.amazon.com/}}.
\newblock


\bibitem[\protect\citeauthoryear{??}{aws}{2019b}]%
        {awslambda}
 \bibinfo{year}{2019}\natexlab{b}.
\newblock \bibinfo{title}{{AWS Lambda}}.
\newblock \bibinfo{howpublished}{\url{https://aws.amazon.com/lambda/}}.
\newblock


\bibitem[\protect\citeauthoryear{??}{lam}{2019}]%
        {lambda-cold}
 \bibinfo{year}{2019}\natexlab{}.
\newblock \bibinfo{title}{{{AWS Lambda Cold Starts}}}.
\newblock
  \bibinfo{howpublished}{\url{https://mikhail.io/serverless/coldstarts/aws/}}.
\newblock


\bibitem[\protect\citeauthoryear{??}{sar}{2019}]%
        {sar}
 \bibinfo{year}{2019}\natexlab{}.
\newblock \bibinfo{title}{{{AWS Serverless Application Repository}}}.
\newblock
  \bibinfo{howpublished}{\url{https://aws.amazon.com/serverless/serverlessrepo/}}.
\newblock


\bibitem[\protect\citeauthoryear{??}{azu}{2019}]%
        {azure-cold}
 \bibinfo{year}{2019}\natexlab{}.
\newblock \bibinfo{title}{{{Azure Functions Cold Start}}}.
\newblock
  \bibinfo{howpublished}{\url{https://mikhail.io/serverless/coldstarts/azure/}}.
\newblock


\bibitem[\protect\citeauthoryear{??}{clo}{2019}]%
        {cloudlab}
 \bibinfo{year}{2019}\natexlab{}.
\newblock \bibinfo{title}{{{Cloudlab}}}.
\newblock \bibinfo{howpublished}{\url{https://cloudlab.us}}.
\newblock


\bibitem[\protect\citeauthoryear{??}{fir}{2019}]%
        {firecracker}
 \bibinfo{year}{2019}\natexlab{}.
\newblock \bibinfo{title}{{{Firecracker MicroVM}}}.
\newblock \bibinfo{howpublished}{\url{https://firecracker-microvm.github.io/}}.
\newblock


\bibitem[\protect\citeauthoryear{??}{goo}{2019}]%
        {googlecloudfunctions}
 \bibinfo{year}{2019}\natexlab{}.
\newblock \bibinfo{title}{{Google Cloud Functions}}.
\newblock \bibinfo{howpublished}{\url{https://cloud.google.com/functions/}}.
\newblock


\bibitem[\protect\citeauthoryear{??}{kub}{2019}]%
        {kubernetes}
 \bibinfo{year}{2019}\natexlab{}.
\newblock \bibinfo{title}{{{Google Container Engine}}}.
\newblock \bibinfo{howpublished}{\url{http://kubernetes.io}}.
\newblock


\bibitem[\protect\citeauthoryear{??}{gra}{2019}]%
        {grafana}
 \bibinfo{year}{2019}\natexlab{}.
\newblock \bibinfo{title}{{Grafana}}.
\newblock \bibinfo{howpublished}{\url{https://grafana.com/}}.
\newblock


\bibitem[\protect\citeauthoryear{??}{pro}{2019a}]%
        {prometheus}
 \bibinfo{year}{2019}\natexlab{a}.
\newblock \bibinfo{title}{{Prometheus}}.
\newblock \bibinfo{howpublished}{\url{https://prometheus.io/}}.
\newblock


\bibitem[\protect\citeauthoryear{??}{pro}{2019b}]%
        {protobuf}
 \bibinfo{year}{2019}\natexlab{b}.
\newblock \bibinfo{title}{{{Protocol Buffers}}}.
\newblock \bibinfo{howpublished}{\url{https://bit.ly/1mISy49}}.
\newblock


\bibitem[\protect\citeauthoryear{??}{ser}{2019}]%
        {serverless-game}
 \bibinfo{year}{2019}\natexlab{}.
\newblock \bibinfo{title}{{{Serverless WarmUp Plugin}}}.
\newblock
  \bibinfo{howpublished}{\url{https://github.com/FidelLimited/serverless-plugin-warmup}}.
\newblock


\bibitem[\protect\citeauthoryear{Akkus, Chen, Rimac, Stein, Satzke, Beck,
  Aditya, and Hilt}{Akkus et~al\mbox{.}}{2018}]%
        {sand}
\bibfield{author}{\bibinfo{person}{Istemi~Ekin Akkus},
  \bibinfo{person}{Ruichuan Chen}, \bibinfo{person}{Ivica Rimac},
  \bibinfo{person}{Manuel Stein}, \bibinfo{person}{Klaus Satzke},
  \bibinfo{person}{Andre Beck}, \bibinfo{person}{Paarijaat Aditya}, {and}
  \bibinfo{person}{Volker Hilt}.} \bibinfo{year}{2018}\natexlab{}.
\newblock \showarticletitle{$\{$SAND$\}$: Towards High-Performance Serverless
  Computing}. In \bibinfo{booktitle}{\emph{2018 $\{$USENIX$\}$ Annual Technical
  Conference ($\{$USENIX$\}$$\{$ATC$\}$ 18)}}. \bibinfo{pages}{923--935}.
\newblock


\bibitem[\protect\citeauthoryear{Ao, Izhikevich, Voelker, and Porter}{Ao
  et~al\mbox{.}}{2018}]%
        {sprocket}
\bibfield{author}{\bibinfo{person}{Lixiang Ao}, \bibinfo{person}{Liz
  Izhikevich}, \bibinfo{person}{Geoffrey~M Voelker}, {and}
  \bibinfo{person}{George Porter}.} \bibinfo{year}{2018}\natexlab{}.
\newblock \showarticletitle{Sprocket: A serverless video processing framework}.
  In \bibinfo{booktitle}{\emph{Proceedings of the ACM Symposium on Cloud
  Computing}}. ACM, \bibinfo{pages}{263--274}.
\newblock


\bibitem[\protect\citeauthoryear{Boucher, Kalia, Andersen, and
  Kaminsky}{Boucher et~al\mbox{.}}{2018}]%
        {micro}
\bibfield{author}{\bibinfo{person}{Sol Boucher}, \bibinfo{person}{Anuj Kalia},
  \bibinfo{person}{David~G Andersen}, {and} \bibinfo{person}{Michael
  Kaminsky}.} \bibinfo{year}{2018}\natexlab{}.
\newblock \showarticletitle{Putting the" Micro" back in microservice}. In
  \bibinfo{booktitle}{\emph{2018 $\{$USENIX$\}$ Annual Technical Conference
  ($\{$USENIX$\}$$\{$ATC$\}$ 18)}}. \bibinfo{pages}{645--650}.
\newblock


\bibitem[\protect\citeauthoryear{Boutin, Ekanayake, Lin, Shi, Zhou, Qian, Wu,
  and Zhou}{Boutin et~al\mbox{.}}{2014}]%
        {apollo}
\bibfield{author}{\bibinfo{person}{Eric Boutin}, \bibinfo{person}{Jaliya
  Ekanayake}, \bibinfo{person}{Wei Lin}, \bibinfo{person}{Bing Shi},
  \bibinfo{person}{Jingren Zhou}, \bibinfo{person}{Zhengping Qian},
  \bibinfo{person}{Ming Wu}, {and} \bibinfo{person}{Lidong Zhou}.}
  \bibinfo{year}{2014}\natexlab{}.
\newblock \showarticletitle{Apollo: Scalable and coordinated scheduling for
  cloud-scale computing}. In \bibinfo{booktitle}{\emph{OSDI}}.
\newblock


\bibitem[\protect\citeauthoryear{Fouladi, Wahby, Shacklett, Balasubramaniam,
  Zeng, Bhalerao, Sivaraman, Porter, and Winstein}{Fouladi
  et~al\mbox{.}}{2017}]%
        {excamera}
\bibfield{author}{\bibinfo{person}{Sadjad Fouladi}, \bibinfo{person}{Riad~S
  Wahby}, \bibinfo{person}{Brennan Shacklett},
  \bibinfo{person}{Karthikeyan~Vasuki Balasubramaniam},
  \bibinfo{person}{William Zeng}, \bibinfo{person}{Rahul Bhalerao},
  \bibinfo{person}{Anirudh Sivaraman}, \bibinfo{person}{George Porter}, {and}
  \bibinfo{person}{Keith Winstein}.} \bibinfo{year}{2017}\natexlab{}.
\newblock \showarticletitle{Encoding, fast and slow: Low-latency video
  processing using thousands of tiny threads}. In
  \bibinfo{booktitle}{\emph{14th $\{$USENIX$\}$ Symposium on Networked Systems
  Design and Implementation ($\{$NSDI$\}$ 17)}}. \bibinfo{pages}{363--376}.
\newblock


\bibitem[\protect\citeauthoryear{Fox, Gribble, Chawathe, Brewer, and
  Gauthier}{Fox et~al\mbox{.}}{1997}]%
        {fox1997cluster}
\bibfield{author}{\bibinfo{person}{Armando Fox}, \bibinfo{person}{Steven~D
  Gribble}, \bibinfo{person}{Yatin Chawathe}, \bibinfo{person}{Eric~A Brewer},
  {and} \bibinfo{person}{Paul Gauthier}.} \bibinfo{year}{1997}\natexlab{}.
\newblock \showarticletitle{Cluster-based scalable network services}. In
  \bibinfo{booktitle}{\emph{ACM SIGOPS operating systems review}},
  Vol.~\bibinfo{volume}{31}. ACM, \bibinfo{pages}{78--91}.
\newblock


\bibitem[\protect\citeauthoryear{Harchol-Balter, Schroeder, Bansal, and
  Agrawal}{Harchol-Balter et~al\mbox{.}}{2003}]%
        {HarcholBalter2003SizebasedST}
\bibfield{author}{\bibinfo{person}{Mor Harchol-Balter}, \bibinfo{person}{Bianca
  Schroeder}, \bibinfo{person}{Nikhil Bansal}, {and} \bibinfo{person}{Mukesh
  Agrawal}.} \bibinfo{year}{2003}\natexlab{}.
\newblock \showarticletitle{Size-based scheduling to improve web performance}.
\newblock \bibinfo{journal}{\emph{ACM Trans. Comput. Syst.}}
  \bibinfo{volume}{21} (\bibinfo{year}{2003}), \bibinfo{pages}{207--233}.
\newblock


\bibitem[\protect\citeauthoryear{Hendrickson, Sturdevant, Harter,
  Venkataramani, Arpaci-Dusseau, and Arpaci-Dusseau}{Hendrickson
  et~al\mbox{.}}{2016}]%
        {openlambda}
\bibfield{author}{\bibinfo{person}{Scott Hendrickson}, \bibinfo{person}{Stephen
  Sturdevant}, \bibinfo{person}{Tyler Harter}, \bibinfo{person}{Venkateshwaran
  Venkataramani}, \bibinfo{person}{Andrea~C. Arpaci-Dusseau}, {and}
  \bibinfo{person}{Remzi~H. Arpaci-Dusseau}.} \bibinfo{year}{2016}\natexlab{}.
\newblock \showarticletitle{Serverless Computation with OpenLambda}. In
  \bibinfo{booktitle}{\emph{HotCloud 16}}.
\newblock


\bibitem[\protect\citeauthoryear{Hindman, Konwinski, Zaharia, Ghodsi, Joseph,
  Katz, Shenker, and Stoica}{Hindman et~al\mbox{.}}{2011}]%
        {mesos}
\bibfield{author}{\bibinfo{person}{B. Hindman}, \bibinfo{person}{A. Konwinski},
  \bibinfo{person}{M. Zaharia}, \bibinfo{person}{A. Ghodsi},
  \bibinfo{person}{A.D. Joseph}, \bibinfo{person}{R. Katz}, \bibinfo{person}{S.
  Shenker}, {and} \bibinfo{person}{I. Stoica}.}
  \bibinfo{year}{2011}\natexlab{}.
\newblock \showarticletitle{{Mesos: A Platform for Fine-Grained Resource
  Sharing in the Data Center}}. In \bibinfo{booktitle}{\emph{NSDI}}.
\newblock


\bibitem[\protect\citeauthoryear{Jonas, Pu, Venkataraman, Stoice, and
  Recht}{Jonas et~al\mbox{.}}{2017}]%
        {pywren}
\bibfield{author}{\bibinfo{person}{Eric Jonas}, \bibinfo{person}{Qifan Pu},
  \bibinfo{person}{Shivaram Venkataraman}, \bibinfo{person}{Ion Stoice}, {and}
  \bibinfo{person}{Benjamin Recht}.} \bibinfo{year}{2017}\natexlab{}.
\newblock \showarticletitle{Occupy the Cloud: Distributed Computing for the
  99\%}. In \bibinfo{booktitle}{\emph{SOCC}}.
\newblock


\bibitem[\protect\citeauthoryear{Jonas, Schleier-Smith, Sreekanti, Tsai,
  Khandelwal, Pu, Shankar, Carreira, Krauth, Yadwadkar, Gonzalez, Popa, Stoica,
  and Patterson}{Jonas et~al\mbox{.}}{2019}]%
        {berkeley_report}
\bibfield{author}{\bibinfo{person}{Eric Jonas}, \bibinfo{person}{Johann
  Schleier-Smith}, \bibinfo{person}{Vikram Sreekanti},
  \bibinfo{person}{Chia-Che Tsai}, \bibinfo{person}{Anurag Khandelwal},
  \bibinfo{person}{Qifan Pu}, \bibinfo{person}{Vaishaal Shankar},
  \bibinfo{person}{Joao Carreira}, \bibinfo{person}{Karl Krauth},
  \bibinfo{person}{Neeraja Yadwadkar}, \bibinfo{person}{Joseph~E. Gonzalez},
  \bibinfo{person}{Raluca~Ada Popa}, \bibinfo{person}{Ion Stoica}, {and}
  \bibinfo{person}{David~A. Patterson}.} \bibinfo{year}{2019}\natexlab{}.
\newblock \bibinfo{title}{Cloud Programming Simplified: A Berkeley View on
  Serverless Computing}.
\newblock
\newblock
\showeprint[arxiv]{cs.OS/1902.03383}


\bibitem[\protect\citeauthoryear{Karanasos, Rao, Curino, Douglas,
  Chaliparambil, Fumarola, Heddaya, Ramakrishnan, and Sakalanaga}{Karanasos
  et~al\mbox{.}}{2015}]%
        {mercury}
\bibfield{author}{\bibinfo{person}{Konstantinos Karanasos},
  \bibinfo{person}{Sriram Rao}, \bibinfo{person}{Carlo Curino},
  \bibinfo{person}{Chris Douglas}, \bibinfo{person}{Kishore Chaliparambil},
  \bibinfo{person}{Giovanni Fumarola}, \bibinfo{person}{Solom Heddaya},
  \bibinfo{person}{Raghu Ramakrishnan}, {and} \bibinfo{person}{Sarvesh
  Sakalanaga}.} \bibinfo{year}{2015}\natexlab{}.
\newblock \showarticletitle{Mercury: Hybrid Centralized and Distributed
  Scheduling in Large Shared Clusters}. In \bibinfo{booktitle}{\emph{USENIX
  ATC}}.
\newblock


\bibitem[\protect\citeauthoryear{Karger, Lehman, Leighton, Panigrahy, Levine,
  and Lewin}{Karger et~al\mbox{.}}{1997}]%
        {consistent}
\bibfield{author}{\bibinfo{person}{David Karger}, \bibinfo{person}{Eric
  Lehman}, \bibinfo{person}{Tom Leighton}, \bibinfo{person}{Rina Panigrahy},
  \bibinfo{person}{Matthew Levine}, {and} \bibinfo{person}{Daniel Lewin}.}
  \bibinfo{year}{1997}\natexlab{}.
\newblock \showarticletitle{Consistent hashing and random trees: Distributed
  caching protocols for relieving hot spots on the World Wide Web}. In
  \bibinfo{booktitle}{\emph{Proceedings of the twenty-ninth annual ACM
  symposium on Theory of computing}}.
\newblock


\bibitem[\protect\citeauthoryear{Kelley}{Kelley}{1961}]%
        {critical-path-def}
\bibfield{author}{\bibinfo{person}{James~E Kelley}.}
  \bibinfo{year}{1961}\natexlab{}.
\newblock \showarticletitle{Critical-path planning and scheduling: Mathematical
  basis}.
\newblock \bibinfo{journal}{\emph{Operations Research}} \bibinfo{volume}{9},
  \bibinfo{number}{3} (\bibinfo{year}{1961}), \bibinfo{pages}{296--320}.
\newblock


\bibitem[\protect\citeauthoryear{Kelley}{Kelley}{1963}]%
        {cpm}
\bibfield{author}{\bibinfo{person}{James~E Kelley}.}
  \bibinfo{year}{1963}\natexlab{}.
\newblock \showarticletitle{The critical-path method: Resources planning and
  scheduling}.
\newblock \bibinfo{journal}{\emph{Industrial scheduling}}  \bibinfo{volume}{13}
  (\bibinfo{year}{1963}), \bibinfo{pages}{347--365}.
\newblock


\bibitem[\protect\citeauthoryear{Klimovic, Wang, Kozyrakis, Stuedi, Pfefferle,
  and Trivedi}{Klimovic et~al\mbox{.}}{2018a}]%
        {pocket-workshop}
\bibfield{author}{\bibinfo{person}{Ana Klimovic}, \bibinfo{person}{Yawen Wang},
  \bibinfo{person}{Christos Kozyrakis}, \bibinfo{person}{Patrick Stuedi},
  \bibinfo{person}{Jonas Pfefferle}, {and} \bibinfo{person}{Animesh Trivedi}.}
  \bibinfo{year}{2018}\natexlab{a}.
\newblock \showarticletitle{Understanding ephemeral storage for serverless
  analytics}. In \bibinfo{booktitle}{\emph{2018 $\{$USENIX$\}$ Annual Technical
  Conference ($\{$USENIX$\}$$\{$ATC$\}$ 18)}}. \bibinfo{pages}{789--794}.
\newblock


\bibitem[\protect\citeauthoryear{Klimovic, Wang, Stuedi, Trivedi, Pfefferle,
  and Kozyrakis}{Klimovic et~al\mbox{.}}{2018b}]%
        {pocket}
\bibfield{author}{\bibinfo{person}{Ana Klimovic}, \bibinfo{person}{Yawen Wang},
  \bibinfo{person}{Patrick Stuedi}, \bibinfo{person}{Animesh Trivedi},
  \bibinfo{person}{Jonas Pfefferle}, {and} \bibinfo{person}{Christos
  Kozyrakis}.} \bibinfo{year}{2018}\natexlab{b}.
\newblock \showarticletitle{Pocket: Elastic ephemeral storage for serverless
  analytics}. In \bibinfo{booktitle}{\emph{13th $\{$USENIX$\}$ Symposium on
  Operating Systems Design and Implementation ($\{$OSDI$\}$ 18)}}.
  \bibinfo{pages}{427--444}.
\newblock


\bibitem[\protect\citeauthoryear{McGrath and Brenner}{McGrath and
  Brenner}{2017}]%
        {4-benchmark}
\bibfield{author}{\bibinfo{person}{Garrett McGrath} {and}
  \bibinfo{person}{Paul~R Brenner}.} \bibinfo{year}{2017}\natexlab{}.
\newblock \showarticletitle{Serverless computing: Design, implementation, and
  performance}. In \bibinfo{booktitle}{\emph{2017 IEEE 37th International
  Conference on Distributed Computing Systems Workshops (ICDCSW)}}. IEEE,
  \bibinfo{pages}{405--410}.
\newblock


\bibitem[\protect\citeauthoryear{Mohan, Sane, Doshi, Edupuganti, Nayak, and
  Sukhomlinov}{Mohan et~al\mbox{.}}{2019}]%
        {agile}
\bibfield{author}{\bibinfo{person}{Anup Mohan}, \bibinfo{person}{Harshad Sane},
  \bibinfo{person}{Kshitij Doshi}, \bibinfo{person}{Saikrishna Edupuganti},
  \bibinfo{person}{Naren Nayak}, {and} \bibinfo{person}{Vadim Sukhomlinov}.}
  \bibinfo{year}{2019}\natexlab{}.
\newblock \showarticletitle{Agile cold starts for scalable serverless}. In
  \bibinfo{booktitle}{\emph{11th $\{$USENIX$\}$ Workshop on Hot Topics in Cloud
  Computing (HotCloud 19)}}.
\newblock


\bibitem[\protect\citeauthoryear{Moritz, Nishihara, Wang, Tumanov, Liaw, Liang,
  Paul, Jordan, and Stoica}{Moritz et~al\mbox{.}}{2017}]%
        {ray}
\bibfield{author}{\bibinfo{person}{Philipp Moritz}, \bibinfo{person}{Robert
  Nishihara}, \bibinfo{person}{Stephanie Wang}, \bibinfo{person}{Alexey
  Tumanov}, \bibinfo{person}{Richard Liaw}, \bibinfo{person}{Eric Liang},
  \bibinfo{person}{William Paul}, \bibinfo{person}{Michael~I. Jordan}, {and}
  \bibinfo{person}{Ion Stoica}.} \bibinfo{year}{2017}\natexlab{}.
\newblock \showarticletitle{Ray: {A} Distributed Framework for Emerging {AI}
  Applications}.
\newblock \bibinfo{journal}{\emph{CoRR}}  \bibinfo{volume}{abs/1712.05889}
  (\bibinfo{year}{2017}).
\newblock
\urldef\tempurl%
\url{http://arxiv.org/abs/1712.05889}
\showURL{%
\tempurl}


\bibitem[\protect\citeauthoryear{Oakes, Yang, Houck, Harter, Arpaci-Dusseau,
  and Arpaci-Dusseau}{Oakes et~al\mbox{.}}{2017}]%
        {pipsqueak}
\bibfield{author}{\bibinfo{person}{Edward Oakes}, \bibinfo{person}{Leon Yang},
  \bibinfo{person}{Kevin Houck}, \bibinfo{person}{Tyler Harter},
  \bibinfo{person}{Andrea~C Arpaci-Dusseau}, {and} \bibinfo{person}{Remzi~H
  Arpaci-Dusseau}.} \bibinfo{year}{2017}\natexlab{}.
\newblock \showarticletitle{Pipsqueak: Lean lambdas with large libraries}. In
  \bibinfo{booktitle}{\emph{2017 IEEE 37th International Conference on
  Distributed Computing Systems Workshops (ICDCSW)}}. IEEE,
  \bibinfo{pages}{395--400}.
\newblock


\bibitem[\protect\citeauthoryear{Oakes, Yang, Zhou, Houck, Harter,
  Arpaci-Dusseau, and Arpaci-Dusseau}{Oakes et~al\mbox{.}}{2018}]%
        {sock}
\bibfield{author}{\bibinfo{person}{Edward Oakes}, \bibinfo{person}{Leon Yang},
  \bibinfo{person}{Dennis Zhou}, \bibinfo{person}{Kevin Houck},
  \bibinfo{person}{Tyler Harter}, \bibinfo{person}{Andrea~C. Arpaci-Dusseau},
  {and} \bibinfo{person}{Remzi~H. Arpaci-Dusseau}.}
  \bibinfo{year}{2018}\natexlab{}.
\newblock \showarticletitle{{SOCK}: Rapid Task Provisioning with
  Serverless-Optimized Containers}. In \bibinfo{booktitle}{\emph{ATC 18}}.
\newblock


\bibitem[\protect\citeauthoryear{Ousterhout, Wendell, Zaharia, and
  Stoica}{Ousterhout et~al\mbox{.}}{2013}]%
        {sparrow}
\bibfield{author}{\bibinfo{person}{Kay Ousterhout}, \bibinfo{person}{Patrick
  Wendell}, \bibinfo{person}{Matei Zaharia}, {and} \bibinfo{person}{Ion
  Stoica}.} \bibinfo{year}{2013}\natexlab{}.
\newblock \showarticletitle{Sparrow: Distributed, low latency scheduling}. In
  \bibinfo{booktitle}{\emph{SOSP}}.
\newblock


\bibitem[\protect\citeauthoryear{Pu, Venkataraman, and Stoica}{Pu
  et~al\mbox{.}}{2019}]%
        {locus}
\bibfield{author}{\bibinfo{person}{Qifan Pu}, \bibinfo{person}{Shivaram
  Venkataraman}, {and} \bibinfo{person}{Ion Stoica}.}
  \bibinfo{year}{2019}\natexlab{}.
\newblock \showarticletitle{Shuffling, fast and slow: scalable analytics on
  serverless infrastructure}. In \bibinfo{booktitle}{\emph{16th $\{$USENIX$\}$
  Symposium on Networked Systems Design and Implementation ($\{$NSDI$\}$ 19)}}.
  \bibinfo{pages}{193--206}.
\newblock


\bibitem[\protect\citeauthoryear{Schrage}{Schrage}{1968}]%
        {schrage1968}
\bibfield{author}{\bibinfo{person}{Linus Schrage}.}
  \bibinfo{year}{1968}\natexlab{}.
\newblock \showarticletitle{A Proof of the Optimality of the Shortest Remaining
  Processing Time Discipline}.
\newblock \bibinfo{journal}{\emph{Operations Research}} \bibinfo{volume}{16},
  \bibinfo{number}{3} (\bibinfo{year}{1968}), \bibinfo{pages}{687--690}.
\newblock


\bibitem[\protect\citeauthoryear{Schwarzkopf, Konwinski, Abd-El-Malek, and
  Wilkes}{Schwarzkopf et~al\mbox{.}}{2013}]%
        {omega}
\bibfield{author}{\bibinfo{person}{Malte Schwarzkopf}, \bibinfo{person}{Andy
  Konwinski}, \bibinfo{person}{Michael Abd-El-Malek}, {and}
  \bibinfo{person}{John Wilkes}.} \bibinfo{year}{2013}\natexlab{}.
\newblock \showarticletitle{Omega: Flexible, scalable schedulers for large
  compute clusters}. In \bibinfo{booktitle}{\emph{EuroSys}}.
\newblock


\bibitem[\protect\citeauthoryear{Shankar, Krauth, Pu, Jonas, Venkataraman,
  Stoica, Recht, and Ragan-Kelley}{Shankar et~al\mbox{.}}{2018}]%
        {numpywren}
\bibfield{author}{\bibinfo{person}{Vaishaal Shankar}, \bibinfo{person}{Karl
  Krauth}, \bibinfo{person}{Qifan Pu}, \bibinfo{person}{Eric Jonas},
  \bibinfo{person}{Shivaram Venkataraman}, \bibinfo{person}{Ion Stoica},
  \bibinfo{person}{Benjamin Recht}, {and} \bibinfo{person}{Jonathan
  Ragan-Kelley}.} \bibinfo{year}{2018}\natexlab{}.
\newblock \showarticletitle{numpywren: serverless linear algebra}.
\newblock \bibinfo{journal}{\emph{arXiv preprint arXiv:1810.09679}}
  (\bibinfo{year}{2018}).
\newblock


\bibitem[\protect\citeauthoryear{Singhvi, Banerjee, Harchol, Akella, Peek, and
  Rydin}{Singhvi et~al\mbox{.}}{2017}]%
        {serverlessNF}
\bibfield{author}{\bibinfo{person}{Arjun Singhvi}, \bibinfo{person}{Sujata
  Banerjee}, \bibinfo{person}{Yotam Harchol}, \bibinfo{person}{Aditya Akella},
  \bibinfo{person}{Mark Peek}, {and} \bibinfo{person}{Pontus Rydin}.}
  \bibinfo{year}{2017}\natexlab{}.
\newblock \showarticletitle{Granular computing and network intensive
  applications: Friends or foes?}. In \bibinfo{booktitle}{\emph{Proceedings of
  the 16th ACM Workshop on Hot Topics in Networks}}. ACM,
  \bibinfo{pages}{157--163}.
\newblock


\bibitem[\protect\citeauthoryear{Vavilapalli, Murthy, Douglas, Agarwal, Konar,
  Evans, Graves, Lowe, Shah, Seth, Saha, Curino, O'Malley, Radia, Reed, and
  Baldeschwieler}{Vavilapalli et~al\mbox{.}}{2013}]%
        {yarn}
\bibfield{author}{\bibinfo{person}{Vinod~Kumar Vavilapalli},
  \bibinfo{person}{Arun~C Murthy}, \bibinfo{person}{Chris Douglas},
  \bibinfo{person}{Sharad Agarwal}, \bibinfo{person}{Mahadev Konar},
  \bibinfo{person}{Robert Evans}, \bibinfo{person}{Thomas Graves},
  \bibinfo{person}{Jason Lowe}, \bibinfo{person}{Hitesh Shah},
  \bibinfo{person}{Siddharth Seth}, \bibinfo{person}{Bikas Saha},
  \bibinfo{person}{Carlo Curino}, \bibinfo{person}{Owen O'Malley},
  \bibinfo{person}{Sanjay Radia}, \bibinfo{person}{Benjamin Reed}, {and}
  \bibinfo{person}{Eric Baldeschwieler}.} \bibinfo{year}{2013}\natexlab{}.
\newblock \showarticletitle{Apache {Hadoop} {YARN}: Yet Another Resource
  Negotiator}. In \bibinfo{booktitle}{\emph{SoCC}}.
\newblock


\bibitem[\protect\citeauthoryear{Verma, Pedrosa, Korupolu, Oppenheimer, Tune,
  and Wilkes}{Verma et~al\mbox{.}}{2015}]%
        {borg}
\bibfield{author}{\bibinfo{person}{Abhishek Verma}, \bibinfo{person}{Luis
  Pedrosa}, \bibinfo{person}{Madhukar Korupolu}, \bibinfo{person}{David
  Oppenheimer}, \bibinfo{person}{Eric Tune}, {and} \bibinfo{person}{John
  Wilkes}.} \bibinfo{year}{2015}\natexlab{}.
\newblock \showarticletitle{Large-scale cluster management at {Google} with
  {Borg}}. In \bibinfo{booktitle}{\emph{EuroSys}}.
\newblock


\bibitem[\protect\citeauthoryear{Wang, Li, Zhang, Ristenpart, and Swift}{Wang
  et~al\mbox{.}}{2018}]%
        {serverless-peek}
\bibfield{author}{\bibinfo{person}{Liang Wang}, \bibinfo{person}{Mengyuan Li},
  \bibinfo{person}{Yinqian Zhang}, \bibinfo{person}{Thomas Ristenpart}, {and}
  \bibinfo{person}{Michael Swift}.} \bibinfo{year}{2018}\natexlab{}.
\newblock \showarticletitle{Peeking Behind the Curtains of Serverless
  Platforms}. In \bibinfo{booktitle}{\emph{ATC 18}}.
\newblock


\end{thebibliography}
	\bibliographystyle{ACM-Reference-Format}
}
\end{document}